\documentclass[10pt, conference, compsocconf]{IEEEtran}
 \IEEEoverridecommandlockouts
\ifCLASSINFOpdf
  \usepackage[pdftex]{graphicx}
\else
\fi

\usepackage{color,xcolor}
\usepackage{xspace}
\usepackage{multirow}
\usepackage{booktabs}
\usepackage{lipsum}

\usepackage{amsmath}
\usepackage{amsfonts}
\usepackage{url}

\usepackage{fancyhdr}

\definecolor{citecolor}{RGB}{34,139,34}
\definecolor{mydarkblue}{rgb}{0,0.08,1}
\definecolor{mydarkgreen}{rgb}{0.02,0.6,0.02}
\definecolor{mydarkred}{rgb}{0.8,0.02,0.02}
\definecolor{mydarkorange}{rgb}{0.40,0.2,0.02}
\definecolor{mypurple}{RGB}{111,0,255}
\definecolor{myred}{rgb}{1.0,0.0,0.0}
\definecolor{mygold}{rgb}{0.75,0.6,0.12}
\definecolor{myblue}{rgb}{0,0.2,0.8}
\definecolor{mydarkgray}{rgb}{0.,0.2,0.2}

\renewcommand\footnotemark{}

\newcommand{\ie}{i.e., }

\newcommand{\spgemm}{SpGEMM\xspace}
\newcommand{\name}{SpArch\xspace}

\newcommand{\belady}{B\'{e}l\'{a}dy\xspace}
\newcommand{\x}{$\times$}

\newcommand{\cusparse}{cuSPARSE\xspace}
\newcommand{\cusp}{CUSP\xspace}
\newcommand{\mult}{\emph{multiply}\xspace}
\newcommand{\merge}{\emph{merge}\xspace}

\newcommand{\outerspace}{{OuterSPACE}\xspace}

\newcommand{\dramaccesssaving}{2.8}

\newcommand{\perfoverouterspace}{4}
\newcommand{\perfovermkl}{19}
\newcommand{\perfovercusparse}{18}
\newcommand{\perfovercusp}{17}
\newcommand{\perfoverarmadillo}{1285}

\newcommand{\eeoverouterspace}{6}
\newcommand{\eeovermkl}{164}
\newcommand{\eeovercusparse}{435}
\newcommand{\eeovercusp}{307}
\newcommand{\eeoverarm}{62}

\begin{document}

\title{\name: Efficient \underline{Arch}itecture for \underline{Sp}arse Matrix Multiplication}

\author{\IEEEauthorblockN{Zhekai Zhang$^{*}$, Hanrui Wang$^{*}$\thanks{\normalsize{$^{*}$Equal Contributions.}}, Song Han}
\IEEEauthorblockA{EECS\\
Massachusetts Institute of Technology\\
Cambridge, MA, US\\
\{zhangzk, hanrui, songhan\}@mit.edu}
\and
\IEEEauthorblockN{William J. Dally}
\IEEEauthorblockA{Electrical Engineering\\
Stanford University / NVIDIA\\
Stanford, CA, US\\
dally@stanford.edu}
}

\maketitle
\thispagestyle{fancy}

\vspace{-100pt}
\begin{abstract}



Generalized Sparse Matrix-Matrix Multiplication (SpGEMM) is a ubiquitous task in various engineering and scientific applications. However, inner product based \spgemm introduces redundant input fetches for mismatched nonzero operands, while outer product based approach~\cite{pal2018outerspace} suffers from poor output locality due to numerous partial product matrices. Inefficiency in the reuse of either inputs or outputs data leads to extensive and expensive DRAM access.
 
To address this problem, this paper proposes an efficient sparse matrix multiplication accelerator architecture, \name, which jointly optimizes the data locality for both input and output matrices. 
We first design a highly parallelized \textit{streaming-based merger} to pipeline the multiply and merge stage of partial matrices so that partial matrices are merged on chip immediately after produced.
We then propose a \textit{condensed matrix representation} that reduces the number of partial matrices by three orders of magnitude and thus reduces DRAM access by 5.4\x. We further develop a \textit{Huffman tree scheduler} to improve the scalability of the merger for larger sparse matrices, which reduces the DRAM access by another 1.8\x. We also resolve the increased input matrix read induced by the new representation using a \textit{row prefetcher} with near-optimal buffer replacement policy, further reducing the DRAM access by 1.5\x.
Evaluated on 20 benchmarks,
\name reduces the total DRAM access by \dramaccesssaving\x \xspace over previous state-of-the-art.
On average, \name achieves \perfoverouterspace\x, \perfovermkl\x, \perfovercusparse\x, \perfovercusp\x \xspace, \perfoverarmadillo\x \xspace speedup and \eeoverouterspace\x, \eeovermkl\x, \eeovercusparse\x, \eeovercusp\x \xspace, \eeoverarm\x \xspace energy savings over \outerspace, MKL, \cusparse, \cusp, and ARM Armadillo, respectively.

\end{abstract}


\IEEEpeerreviewmaketitle

\section{Introduction}
\label{sec:introduction}
\begin{figure}[t]
    \centering
    \includegraphics[width=\columnwidth]{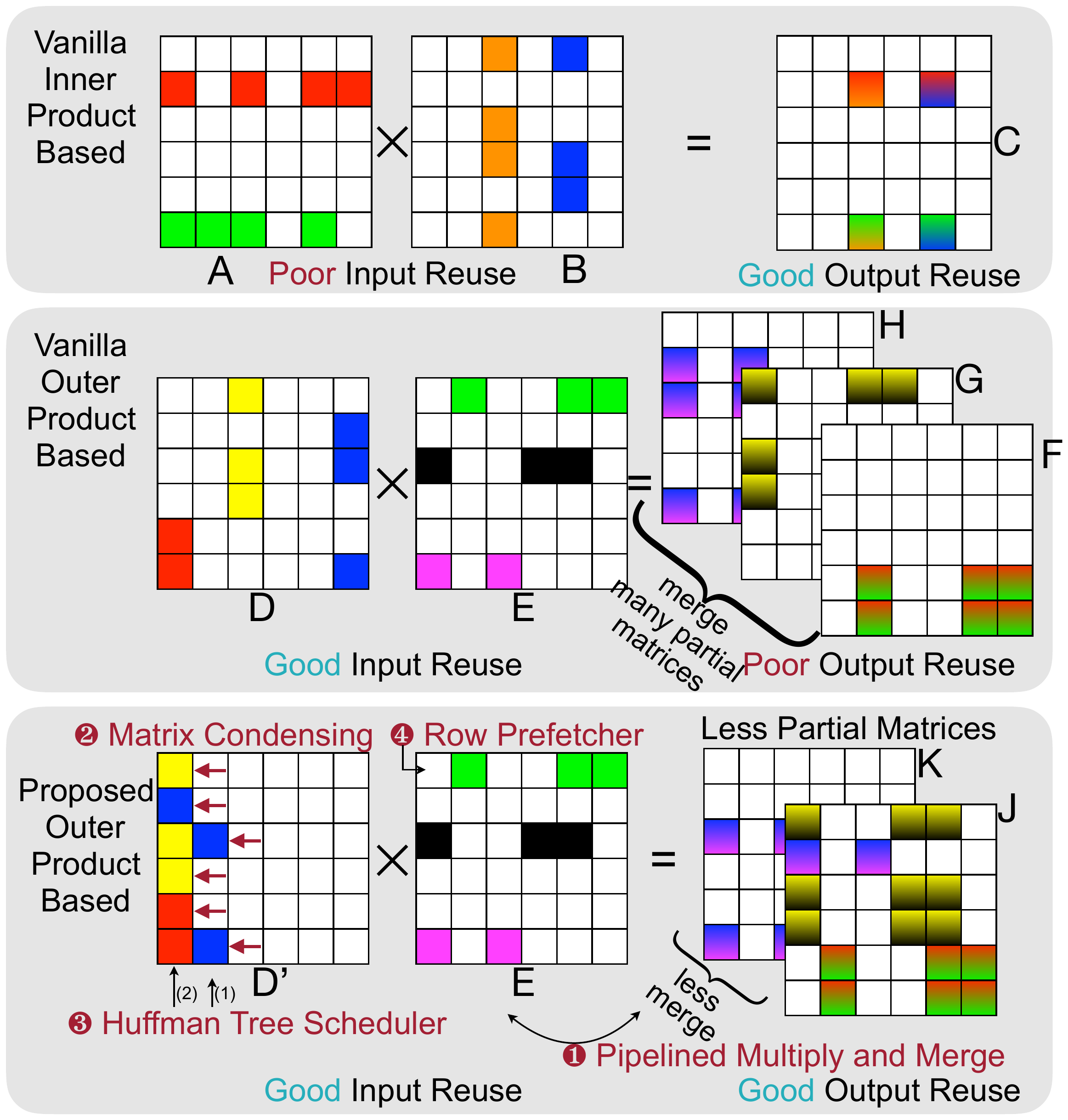}
    \vspace{-10pt}

    \caption{Proposed outer product based \spgemm architecture jointly optimizes input and output data reuse. It achieves good output reuse with pipelined multiply and merge, matrix condensing, Huffman tree scheduler, good input reuse with row prefetcher. 
    }
    \vspace{-15pt}

    \label{fig:teaser}
\end{figure}

\begin{figure*}[t]
    \centering
    \includegraphics[width=\textwidth]{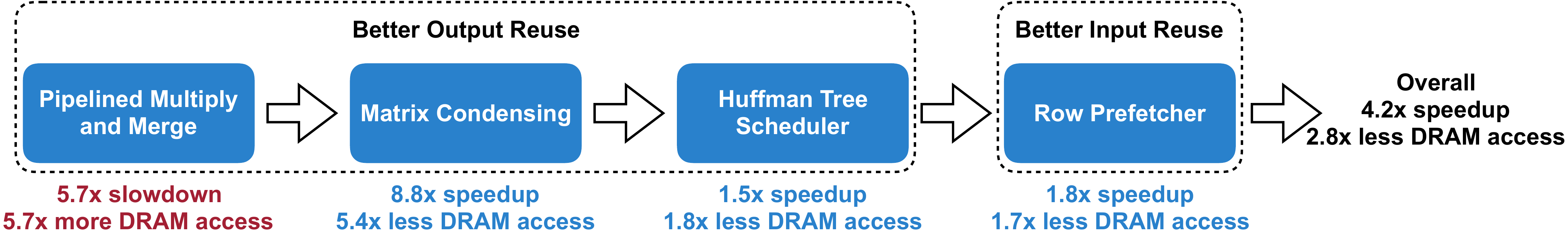}
    \vspace{-15pt}
    \caption{Four innovations in \name. The first slowdown is because of the excessive memory access of partially merged results when the number of partial matrices is much larger than merge tree parallelism. But it enables the next three optimizations.
    We save DRAM bandwidth, achieve higher memory utilization and speedup.}
    \vspace{-10pt}
    \label{fig:4step}
\end{figure*}
Generalized sparse matrix-matrix multiplication (SpG- EMM) is the key computing kernel for many algorithms such as compressed deep neural networks~\cite{han2015deep, han2015learning, wen2016learning, he2018amc}, triangle counting~\cite{azad2015parallel}, Markov clustering~\cite{van2000graph}, searching algorithms~\cite{gilbert2008unified, gilbert2006high}, and matching algorithms~\cite{rabin1989maximum}. It is also ubiquitous in scientific and engineering applications such as grammar parsing~\cite{penn2006efficient}, chemical molecule dynamics~\cite{itoh1995order}, color intersection search~\cite{chan2010more}, linear solver~\cite{yamazaki2010techniques} and many other applications~\cite{bell2012exposing, karypis1994parallel, badia2014highly, wang2018learning, wang2020tts, NIPS20198519}.

However, the performance of \spgemm is memory bounded on the traditional general-purpose computing platforms such as CPUs and GPUs, because of the irregular memory access pattern and poor locality caused by low-density matrices~\cite{goumas2008understanding, cong2018understanding, matam2012sparse}. For instance, the density of Twitter's~\cite{twitter_users} adjacency matrix is as low as 0.000214\%.

As Moore's Law~\cite{moore1965cramming} is slowing down,
domain-specific architecture~\cite{hennessy2018new} becomes promising to improve performance and energy efficiency. \outerspace \cite{pal2018outerspace} proposed \textit{outer product} based \spgemm, which has perfect input reuse compared to \textit{inner product} based method. 
However, outer product has poor output reuse because it produces a considerable amount of partial matrices ($\mathbb{F, G, H}$ in Figure~\ref{fig:teaser}). Those partial matrices need to be stored to DRAM before merging, which incurs extensive DRAM access and cancels the benefits of good input reuse. As a result,
the performance of \outerspace is only 10.4\% of the theoretical peak. 

In this paper, we propose \name\footnote{SpArch is the homophone of word \emph{spark}, meaning striking together two hard surfaces such as stone or metal, which we consider as analogous to \emph{matrix-condensing}, a critical technique in our architecture.}, a domain-specific accelerator to \emph{jointly optimize} input and output data reuse. We obtain input reuse by outer product, and output reuse by on-chip partial matrix merging.

To achieve this, we design a highly parallelized merger to pipeline the two computing stages, \mult and \merge. The multiply stage produces partial matrices, and the \merge stage merges partial matrices into final results. For large matrices, however, the number of partial matrices exceeds the merger's parallelism.
Merging only a part of partial matrices at a time with multiple rounds leads to an increased amount of memory access for the partially merged results, which neutralizes the performance gain of pipelined multiple and merge, making DRAM access even larger (see the first step of Figure~\ref{fig:4step}). However, it is a prerequisite for the next three optimizations.
Therefore, we propose a condensed matrix representation for the first input matrix, where all non-zero elements are pushed to the left, forming much denser columns. As shown in Figure~\ref{fig:teaser}, we condense the first input matrix $\mathbb{D}$ to condensed $\mathbb{D'}$.  The condensed matrix $\mathbb{D'}$ is loaded by condensed column. Implementation-wise, this is equivalent to storing the matrix in CSR format and fetching the elements with the same index for all rows. The second input matrix $\mathbb{E}$ is stored in CSR format. The original left matrix $\mathbb{D}$ has three columns and produces three partial matrices. After condensing, $\mathbb{D'}$ only has two columns and only two partial matrices. In real-world benchmarks, we save the number of partial matrices by three orders of magnitude. 

Unfortunately, the condensed representation can still produce a larger amount of partial matrices than the merger's parallelism. The merge order impacts the amount of DRAM access: partial matrices merged \textit{early} have to be load/stored to DRAM for \textit{every} future merges. Therefore, we should merge matrices with \textit{less} non-zeros first. We design a \emph{Huffman tree scheduler} to decide the near-optimal merge order. We represent each column of a sparse matrix as a leaf node. The weight of the node is the number of non-zeros in that column. For very sparse matrices, the number of non-zeros after merge is approximately the sum of the number of non-zeros before merge. Therefore, the weight of a parent node is the sum of the children's weights, exactly following the convention of a Huffman tree. \textbf{A Huffman tree minimizes the sum of all nodes' weights. We apply it to minimize the total DRAM traffic.}
Scheduled by a Huffman tree, we merge matrices with fewer non-zeros first and larger ones later (example in Figure~\ref{fig:huff}). 
The three techniques discussed above save output DRAM access on 20 benchmarks by $1/5.7\times5.4\times1.8 = 1.7\times$ compared to \outerspace (Figure~\ref{fig:4step}).

However, in \emph{matrix condensing} step, we ruin the perfect reuse of the second matrix ($\mathbb{E}$ in Figure~\ref{fig:teaser}) because now one condensed column of $\mathbb{D'}$ need all the three (green, black and pink) rows of $\mathbb{E}$. In contrast, one column of $\mathbb{D}$ only needs one row of $\mathbb{E}$. Therefore, we propose a \emph{row prefetcher} to prefetch rows of the second matrix $\mathbb{E}$ and store to a row buffer. The buffer replacement policy is near-optimal because we look at the sequence of the rows we need ahead of time while streaming in the first matrix, and thus we can replace the line with farthest next use. 
The row buffer can achieve a 62\% hit rate, thus reducing DRAM access of the second matrix by 2.6\x, largely recovering the input reuse. With the four techniques together, we reduce DRAM access by 2.8\x \ over \outerspace.

In summary, \name jointly optimizes input and output reuse of \spgemm and makes five contributions:
\begin{itemize}
    \item Pipeline the multiply stage and the merge stage with a comparator array-based highly parallelized \textit{merger}.
    
    \item Use \textit{matrix-condensing} to condense the first input matrix, reducing the number of partial matrices by three orders of magnitude.
    
    \item A \textit{Huffman tree scheduler} that provides a near-optimal merge order of partial matrices and reduces the DRAM traffic.
    
    \item A \textit{row prefetcher} that achieves near-optimal buffer replacement policy for the second input matrix and resolves the increased DRAM read induced by matrix-condensing.

    \item We evaluate \name on real-world datasets from SuiteSparse~\cite{datasetuf}, SNAP~\cite{datasetstan} and rMAT~\cite{murphy2010introducing}, achieving \perfoverouterspace\x, \perfovermkl\x, \perfovercusparse\x, \perfovercusp\x, \perfoverarmadillo\x \xspace speedup, and \eeoverouterspace\x, \eeovermkl\x, \eeovercusparse\x, \eeovercusp\x,  \xspace \eeoverarm\x \xspace energy saving over \outerspace, MKL, \cusparse, \cusp and ARM Armadillo.

\end{itemize}

\section{Proposed Architecture}
\label{sec:outerproduct}

\subsection{Comparator Array based Merger}
\label{sec:merger}
Previous state-of-the-art SpGEMM accelerator~\cite{pal2018outerspace}  processes multiply and merge stages separately, which need to store all partial matrices in DRAM. A better way is to pipeline the multiply and merge stages and perform an on-chip merge. 
The partial matrix is represented in COO format with [row index, column index, value]. It is sorted by row index then column index. The merger combines multiple sorted arrays into a new sorted array. A simple merger  puts each array into a FIFO and selects the smallest elements from the top of all FIFOs. This method suffers from low parallelism and thus cannot fully utilize the DRAM bandwidth. Therefore, we design a parallel merge unit. 

\subsubsection{Parallel Merge Unit}
\begin{figure}[t]
    \centering
    \includegraphics[width=\columnwidth]{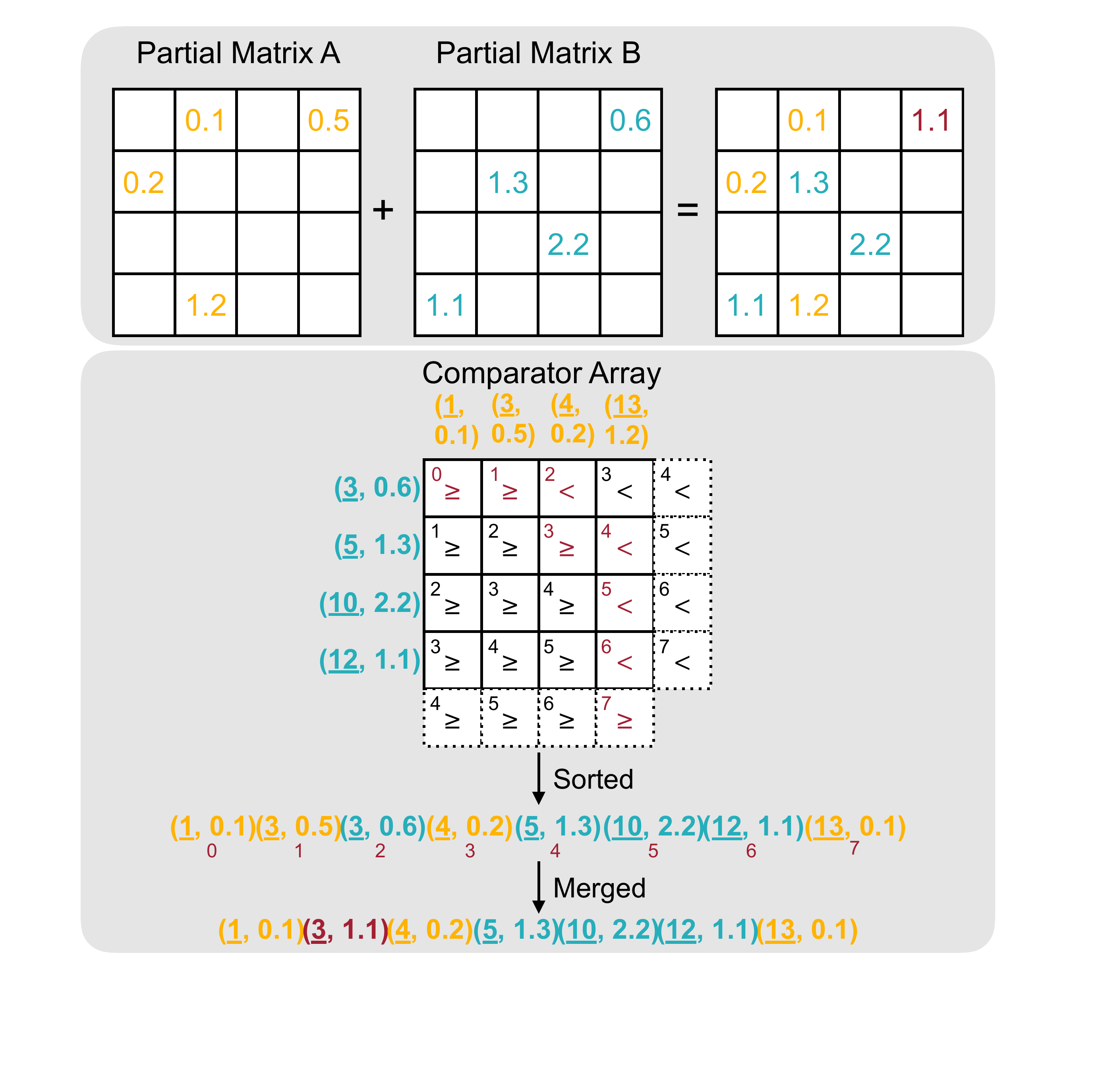}
    \vspace{-10pt}
    \caption{Comparator Array Based Merger. Each diagonal group outputs the data on the boundary of "$\geq$" and "$<$" to form a sorted array. The adder and zero eliminator will further merge the duplicated elements.}
    \label{fig:comparray}
    \vspace{-10pt}
\end{figure}
\label{sec:plainmerger}

A na\"ive merger walks two pointers over two arrays and compares the corresponding elements. The pointer with the smaller element will move forward. This only outputs one element per cycle. In order to increase the parallelism, we replace the pointer by a sliding window of size $N$, all elements in the window A will be compared to all elements in window B by an array of comparators. After the comparison, we move one of the window forward by $N$, improving the throughput by $N$ times.

A comparator array is the core of our merge unit.
As in Figure~\ref{fig:comparray}, the merger contains 4\x4 comparators. The two input matrices are stored in the format of [coordinate, value]. We use the comparator array to compare coordinates of all non-zero elements in partial matrix A and those of partial matrix B. If A$<$B, then the entry is '$<$', otherwise it is '$\ge$'. We pad one dummy column of '$<$' to the right  and one dummy row of '$\ge$' to the bottom. Then we detect a boundary between the '$\ge$' and '$<$' tiles.   The boundary is defined as below: 1. The left-top corner tile is a boundary. 2. The '$\ge$' tiles in the first row are boundaries. 3. If a tile is '$\ge$' and its top neighbor is '$<$', it is a boundary. 4. If a tile is '$<$' and its left neighbor is '$\ge$', it is a boundary. With the rules, we mark the boundary tiles with red in Figure~\ref{fig:comparray}. We further divide all the tiles diagonally into eight groups. The tiles in the same group have the same group index marked at the top-left of each tile. Each group will have \emph{one and only one} output. The output of one boundary tile is the smaller coordinate and corresponding value in its two inputs. For example, for the left-top tile, $1 < 3$, therefore (\underline{1}, 0.1) is the output of that tile, also the output of group 0. 

\begin{figure}[t]
    \centering
    \includegraphics[width=\columnwidth]{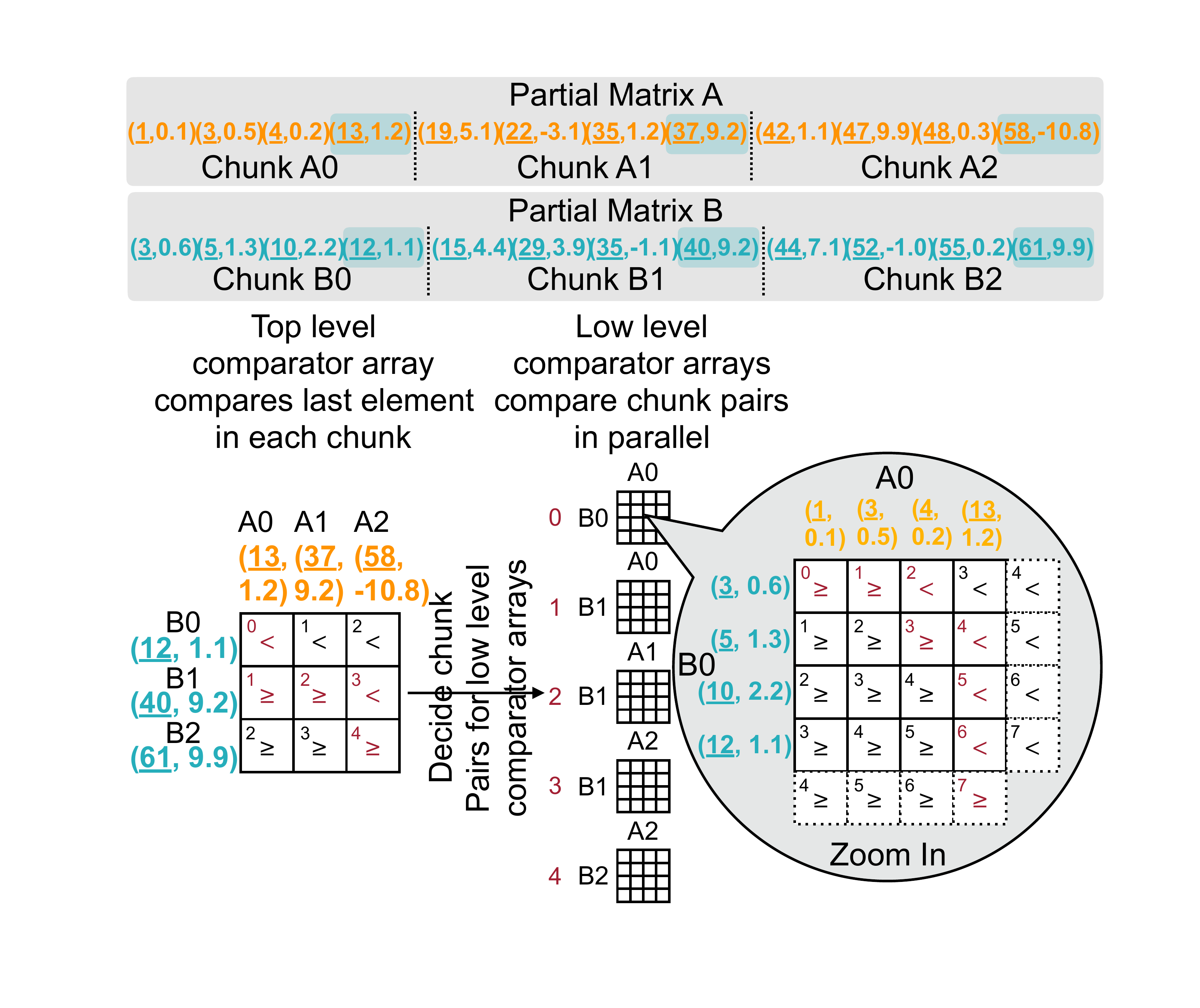}
    \vspace{-15pt}
    \caption{Hierarchical Comparator Array. We break the input array into multiple chunks and use a top level comparator array to decide which pairs of chunks will be fed to low level comparator arrays.}
    \vspace{-15pt}
    \label{fig:comphier}
\end{figure}

The output of the whole comparator array is the sorted results of two input arrays. Let the left input array as $a$ and top array as $b$. The correctness is based on the fact that a boundary tile $(i,j)$ in $k$-th group always has '$<$' above and '$\ge$' on the left. If it is '$\ge$', it will output the top corresponding $b_j$. The '$<$' above indicates $a_0$ to $a_{i-1}$ from left array $a$ are smaller than $b_j$. Plus $b_0$ to $b_{j-1}$, there are exactly $i+j=k$ items smaller than it. Thus the output of that tile will be the $k$-th item in the merged result. The case of '$<$' is similar to '$\ge$'. 

All the results are generated in one clock cycle because there is \emph{no data dependency} between inputs. The 4\x4 comparator array can process arbitrary length of inputs. In each clock cycle, it merges eight inputs, four each from two matrices. Then we shift-in following inputs and replace the previous ones, so new merged results can be produced in each cycle. 

\subsubsection{Hierarchical Parallel Merge Unit}
\begin{figure}
    \centering
    \includegraphics[width=\columnwidth]{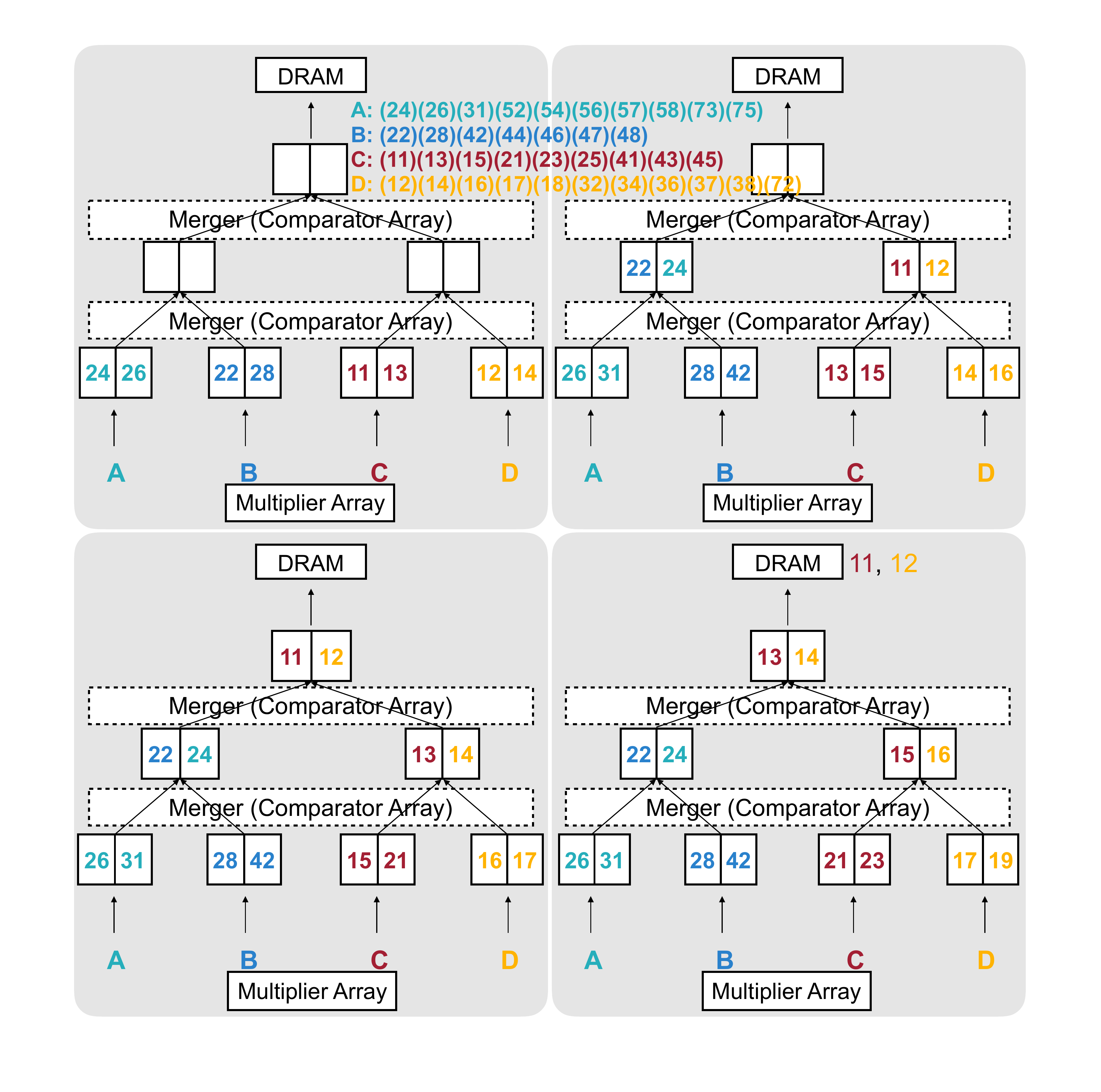}
    \vspace{-15pt}
    \caption{A merge tree which merges four partial product matrices. We only show coordinates of elements here. 
    } 
    \vspace{-15pt}
    \label{fig:mergetree}
\end{figure}
To increase the merger parallelism, we can increase the size of the comparator array, such as from 4\x4 to 12\x12. Nevertheless, the number of comparators is $\mathcal{O}(n^2)$, where $n$ is the side length of the comparator array, thus consuming a large number of hardware resources. Therefore, we further propose a hierarchical merger in Figure~\ref{fig:comphier}. We use two levels of comparator arrays to reduce the total number of comparators.

The hierarchical merge unit contains high-level and low-level comparator arrays. We divide the input into chunks of 4, as in Figure~\ref{fig:comphier} top. The length of each chunk equals to the input length of the low level comparator array (4 in the example). The top level array is used to decide which chunks need to be compared by low level arrays.
The intuition is that if the largest element of chunk A is still smaller than chunk B, then we can skip the comparison of the two chunks.
We use the top level comparator array to compare the last element in each chunk and get the boundary tiles (marked in red). Since each chunk is already sorted, the last element is the largest. The boundary tiles indicate the chunk pairs for the low level comparator arrays. For example, in Figure~\ref{fig:comphier}, the right-bottom tile is a boundary tile, so the A2 and B2 chunk is a chunk pair. We also divide the top level comparator array into diagonal groups, and each group will generate \emph{one and only one} chunk pair. Then we use five low level comparator arrays to process the chunk pairs in parallel.

Output of each low level comparator array is limited by a \emph{min-max bound} to avoid element duplication. Results not in the bounds are set to 0.
If the top level boundary tile has one left/top boundary tile neighbor, then the min bound is the first element of the top/left input chunk. The min bound of the first low level array is the smallest element. The max bound of each low level array is the min bound of the next one. The upper bound of the last low level array is $+\infty$.

Then the outputs of each low level array are concatenated together to get the overall output. In this way, we save the low level comparators in non-boundary tiles in the high level array. Mathematically, the number of comparators is reduced to $\mathcal{O}(n^{\frac{4}{3}})$. If we choose a $n^{\frac{2}{3}}\times n^{\frac{2}{3}}$ top level comparator array and $n^{\frac{1}{3}}\times n^{\frac{1}{3}}$ low level comparators, we can process $n$ elements at a time and uses only $(2n^{\frac{2}{3}}-1)*(n^{\frac{1}{3}})^2+(n^{\frac{2}{3}})^2=\mathcal{O}(n^{\frac{4}{3}})$ comparators. 

\subsubsection{Merge Tree}

Using the hierarchical merger, we get a highly parallelized binary streaming merger that can merge up to 16 elements in each cycle. It merges two arrays into one array. In order to merge more arrays into one array, we stack multiple binary mergers and form a merge tree.
As shown in Figure~\ref{fig:mergetree}, the merge tree is a full binary tree that each node represents a FIFO on the hardware. Input arrays are fed to the leaf nodes, and the output array is collected from the root node. 
A binary merger merges the elements from the child FIFO and stores the results to the parent FIFO.
The throughput of the whole tree is bounded by the root node, which has only one merger. Therefore, each layer \emph{shares one merger to balance the throughput}. 

Figure~\ref{fig:mergetree} illustrates a merge tree of 2 layers, and each layer is equipped with a 1\x1 merger. We only show coordinates of elements here. Four arrays A, B, C, D are to be merged. They are loaded into the leaf FIFOs. In the first 4 cycles, the lower merger merges data from the leaf FIFOs and store the results to the middle FIFOs. Then the upper merger notices there is enough data in the middle FIFOs, so it merges them and stores to the topmost FIFO. When the root FIFO is full, data is written to DRAM. The merge tree continues to working streamingly until all data fed to the lowest FIFO reach the root, and four arrays are merged fully on-chip. 

\subsubsection{Adders and Zero Eliminator}
\begin{figure}
    \centering
    \includegraphics[width=1\columnwidth]{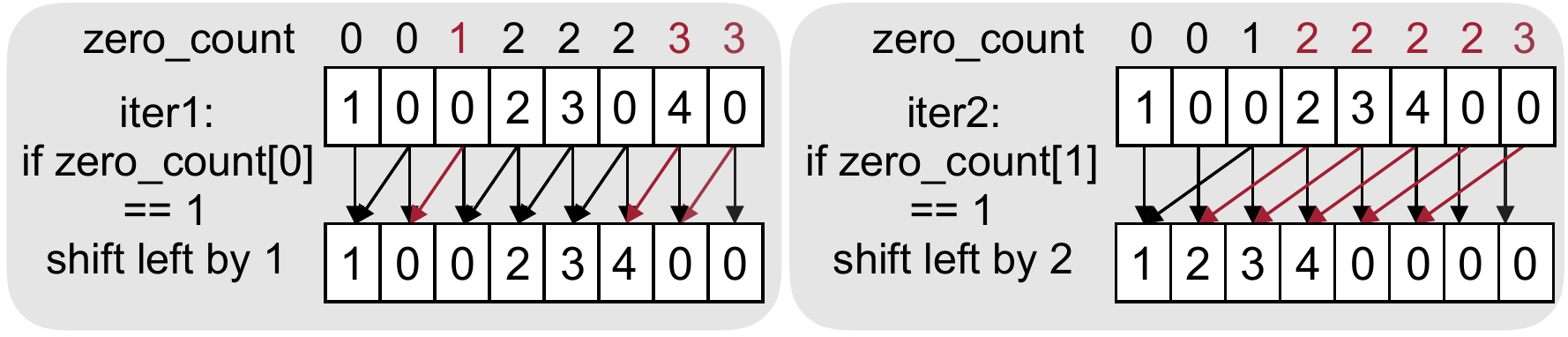}
    \caption{Zero Eliminator. Bit 0 of zero\_count is checked to determine whether to shift by 1 in the first layer. Bit 1 of $zero\_count$ is checked to determine whether to shift by 2 in the second layer. We need $\log{N}$-cycle-latency to process input of length N.}
    \vspace{-10pt}
    \label{fig:ze}
\end{figure}
\begin{figure}
    \centering
    \includegraphics[width=\columnwidth]{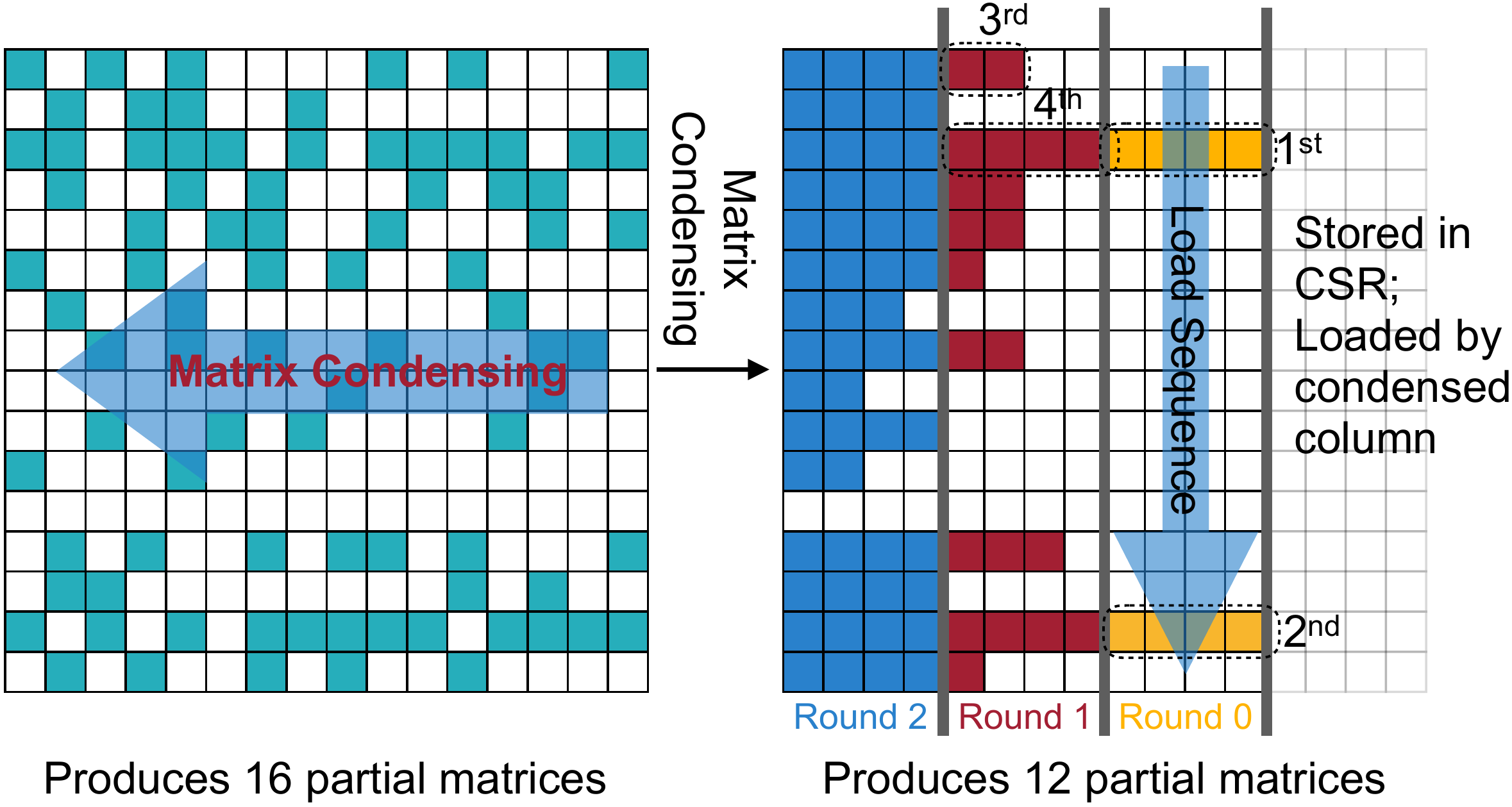}
    \caption{Matrix Condensing. Condense the sparse matrix to the left, reducing the number of columns, thus reducing the number of partial matrices. It can be stored naturally using the CSR format.} 
    \vspace{-10pt}
    \label{fig:matcon}
\end{figure}
\begin{figure*}[t]
    \centering
    \includegraphics[width=\textwidth]{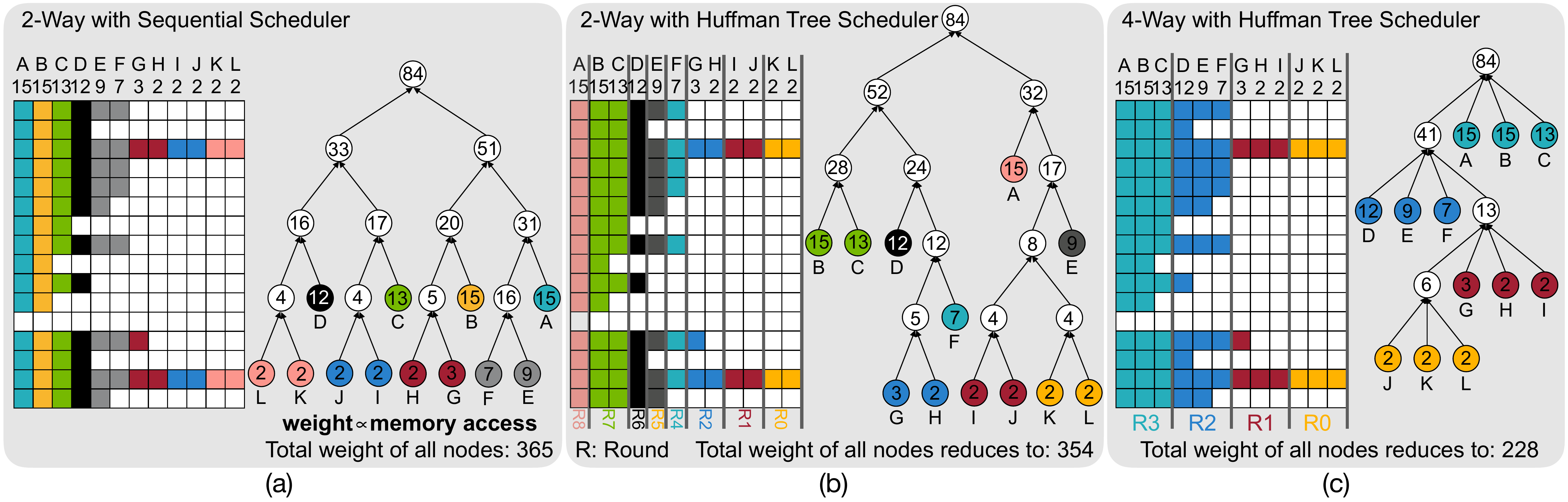}
    \vspace{-15pt}
    \caption{Huffman tree scheduler gives near optimal order of partial matrices merging. The total memory access is minimized. The weights of nodes represent the number of non-zeros in the partial matrix.  The total  DRAM access is proportional to the sum of nodes' weights.
    }
    \vspace{-15pt}
    \label{fig:huff}
\end{figure*}
The merger stated above only merges the elements and leaves alone same-location elements, \ie elements with the same row and column index. However, in SpGEMM, we need to perform add operation on those elements. Therefore, we connect a slice of adders right after the merger, and it will add adjacent same-location elements and set one of the elements to zero. Then we use a Zero Eliminator to compress these zeroes and output the dense results.
The Zero Eliminator consists of two parts. The first part is a prefix sum module that computes the number of zeroes ($zero\_count$) before each element. The second part is a modified $\log N$ layer shifter. Each layer contains $N$ MUXs that shift the input array and $zero\_count$ by 1,2,4,... positions. Unlike a traditional shifter in which MUXs share a common control signal, the MUXs in Zero Eliminator are controlled by the $zero\_count$ signal of each element. Figure~\ref{fig:ze} shows an example of the Zero Eliminator.

\subsection{Matrix Condensing}
\label{sec:condensing}

Due to the limitation of hardware resources, we can only afford to merge 64 arrays on-chip. However, the intermediate results generated by the multiplier can be as large as 10,000 to 1,000,000, which requires multiple rounds of 64-way merging and still consumes a considerable amount of DRAM bandwidth. Therefore, we propose matrix condensing to merge multiple columns into a single column, which reduces the number of partial matrices to 100 to 1,000 in our benchmarks.

The motivation of matrix condensing is that if we have two columns $a_1$, $a_2$ from the left matrix that have no elements sharing the same row index, we can merge them (i.e., combine them into a new array sorted by the row index) while keeping the original column index. When we process the merged column in multiply phase, we multiply each element by its corresponding row (the same as its original column index) in the second input matrix $B$. The result is the same as the merge result of $a_1 \times B$ and $a_2 \times B$. We use a cheap merge of the left matrix to replace an expensive merge of the much longer multiplied results and reduce the number of partial matrices.

Furthermore, since exchanging elements between different columns does not affect the final result, we condense all elements in a row to the leftmost column. In this way, the number of columns of the condensed left matrix is far less than the original one. The number of partial product matrices is equivalent to the number of condensed columns, which is reduced by three orders of magnitude.

Figure~\ref{fig:matcon} shows a matrix in condensed format. The number of columns is reduced from 16 to 12, which equals the length of the longest row in the original matrix. In real-world datasets, we can reduce it from 100,000 to 100\textasciitilde 1,000, which is much closer to the size of the merge tree.

We store the left matrix in CSR format. The elements in CSR directly map to those in condensed format: the $i^{\mathrm{th}}$ element in a CSR row is just in the $i^{\mathrm{th}}$ column of condensed format. 
CSR format and our condensed format are two different views of the same data. In the view of DRAM, the matrix is loaded by the rows, but in the view of a port of the merge tree, the matrix is loaded by condensed columns. This is achieved by the data loader, which dispatches data to different ports according to its condensed column.
As in Figure~\ref{fig:matcon}, if the merger has parallelism of 4, we load four condensed columns together. The dash frames show the load sequence. After loading an element from the left matrix, we can use its original column index to fetch a row from the right matrix and fed them to the multipliers. The multiplied results will be fed to the ports whose index equals to the condensed column index (not the original column index). 

\subsection{Huffman Tree Scheduler}
\label{sec:huffman}

The merge unit can merge 64 matrices on-chip. After matrix condensing, the number of partial matrices can still exceed 64, which needs to be written to DRAM and merged later. The order of the merge matters: the earlier a matrix is merged, the more rounds of DRAM read and write it needs. Intuitively, we should merge sparser partial products matrices first, since they have less number of non-zeros; even in-and-out from DRAM for many times, the access number is smaller. We should leave denser matrices merged later.
To wisely choose the order, we use a Huffman tree scheduler to minimize memory access during the whole task.

Our scheduler abstracts the entire merge process as a tree. In Figure~\ref{fig:huff}, we show the 2-way and 4-way Huffman tree schedulers when the merger can merge 2, and 4 matrices at the same time. We also show a 2-way sequential scheduler without the Huffman tree scheduler for comparison.

The leaf nodes in the tree represent the initial multiplied results of a column in the left matrix with the right matrix. The internal nodes represent the partially merged results. The weights of the nodes represent the size of partially merged results or the size of the initial multiplied results.
For internal nodes, we estimate its weight by adding up the weights of its children because the matrix is very sparse, and the additions during the merge stage are relatively rare.
The arrow points from nodes to their parent node represent a round of multiply-and-merge operation.

The memory access amount of all partially merged results equals to the sum of all internal node weights. Since the root node and leaf node weights do not rely on the tree shape, the optimization goal is to minimize the total weights of all nodes, which also equals the sum of the weights of leaf nodes multiplied by their depths in the tree.

A k-ary Huffman Tree is proven to be the optimal solution to minimize the sum of all nodes. It puts larger leaf nodes nearer to the root and smaller nodes farther, so the accumulative products of weights and depths can be minimized. This is exactly what we need to minimize the total DRAM traffic.
In each round of the Huffman Tree construction, we select k un-merged nodes with \textit{minimal} weights and merge them into an internal node except the first round. In the first round, the number of nodes that need to choose can be calculated with Formula~\ref{form:nummat}. This guarantees that the last round always merges k nodes, so the root node of the tree is \emph{always full}.
\begin{equation}
\label{form:nummat}
k_{init} = (num_{condensed\_col} - 2) mod (num_{merger\_way}-1) + 2
\end{equation}

In the example (Figure~\ref{fig:huff}), a 2-way Huffman scheduler reduced the total weights from 365 to 354. A 4-way Huffman scheduler further reduces it to 228. The weight is proportional to the DRAM traffic. Therefore total weight is reduced when the parallelism of the merger increases. However, high parallelism usually comes at the cost of high power consumption and large area. We choose a 64-way Huffman tree scheduler, and will discuss the trade-offs in Section~\ref{sec:experimental}.

In our real implementation, the Huffman tree is built on the fly with a priority queue. The implementation reflects the algorithm of building a Huffman tree. We firstly add the weights of leaf nodes to the queue and sort them. For a $m-$way merger, in each iteration, the first $m$ partial matrices are merged, and the weight of the merged matrix is added to the queue, and then we sort the queue again.

\subsection{Row Prefetcher}
\label{sec:prefetcher}

\begin{figure*}[t]
    \centering
    \includegraphics[width=\textwidth]{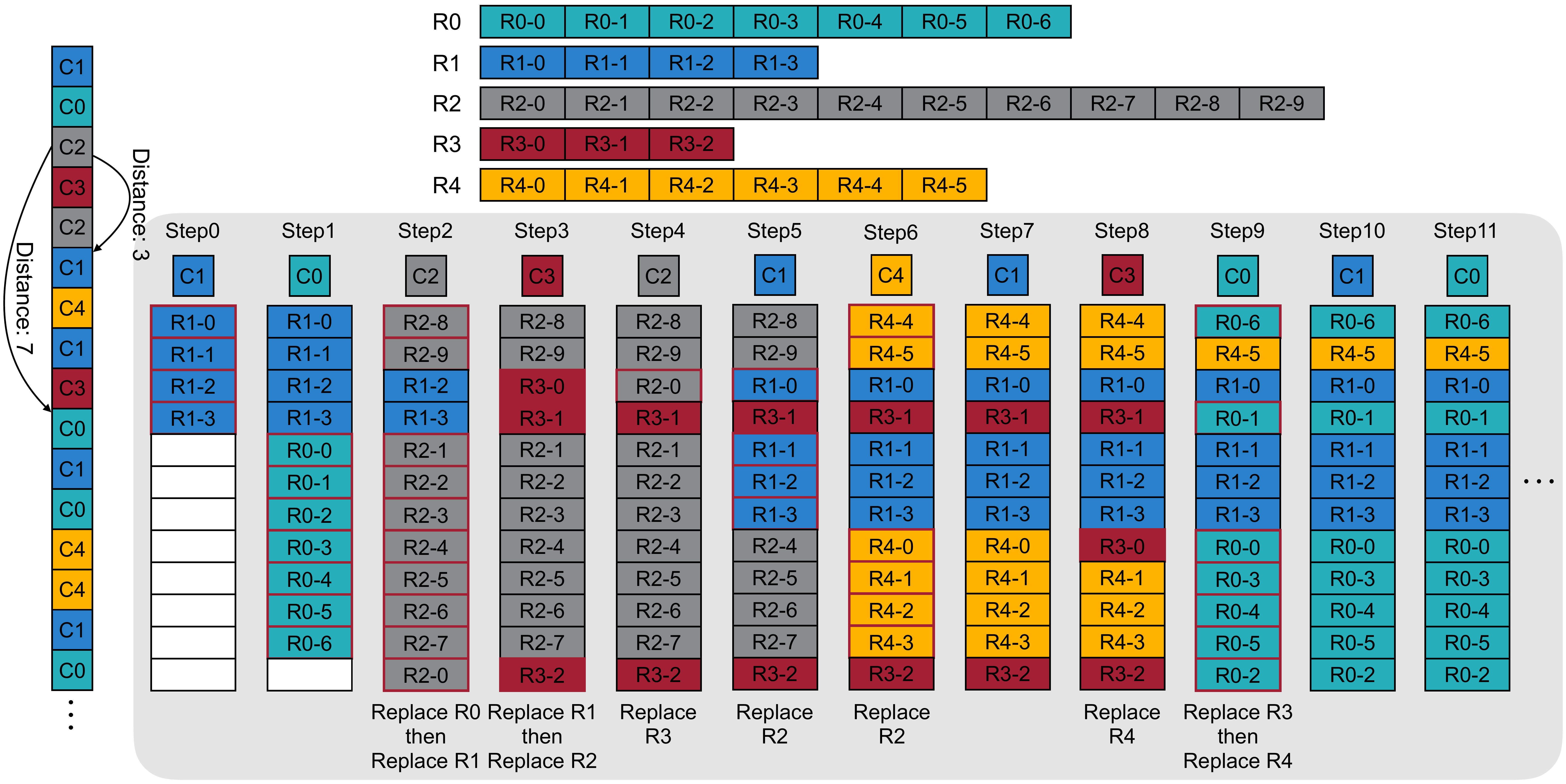}
    \vspace{-10pt}
    \caption{The row prefetcher and the buffer replacement policy. The prefetcher fetches the rows of the second matrix in advance to hide the DRAM latency. The buffer lines are replaced according to the  pre-calculated sequence of required rows thus can achieve near \belady optimal.}
    \vspace{-15pt}
    \label{fig:prefetcher}
\end{figure*}
\begin{figure}[t]
    \centering
    \includegraphics[width=\columnwidth]{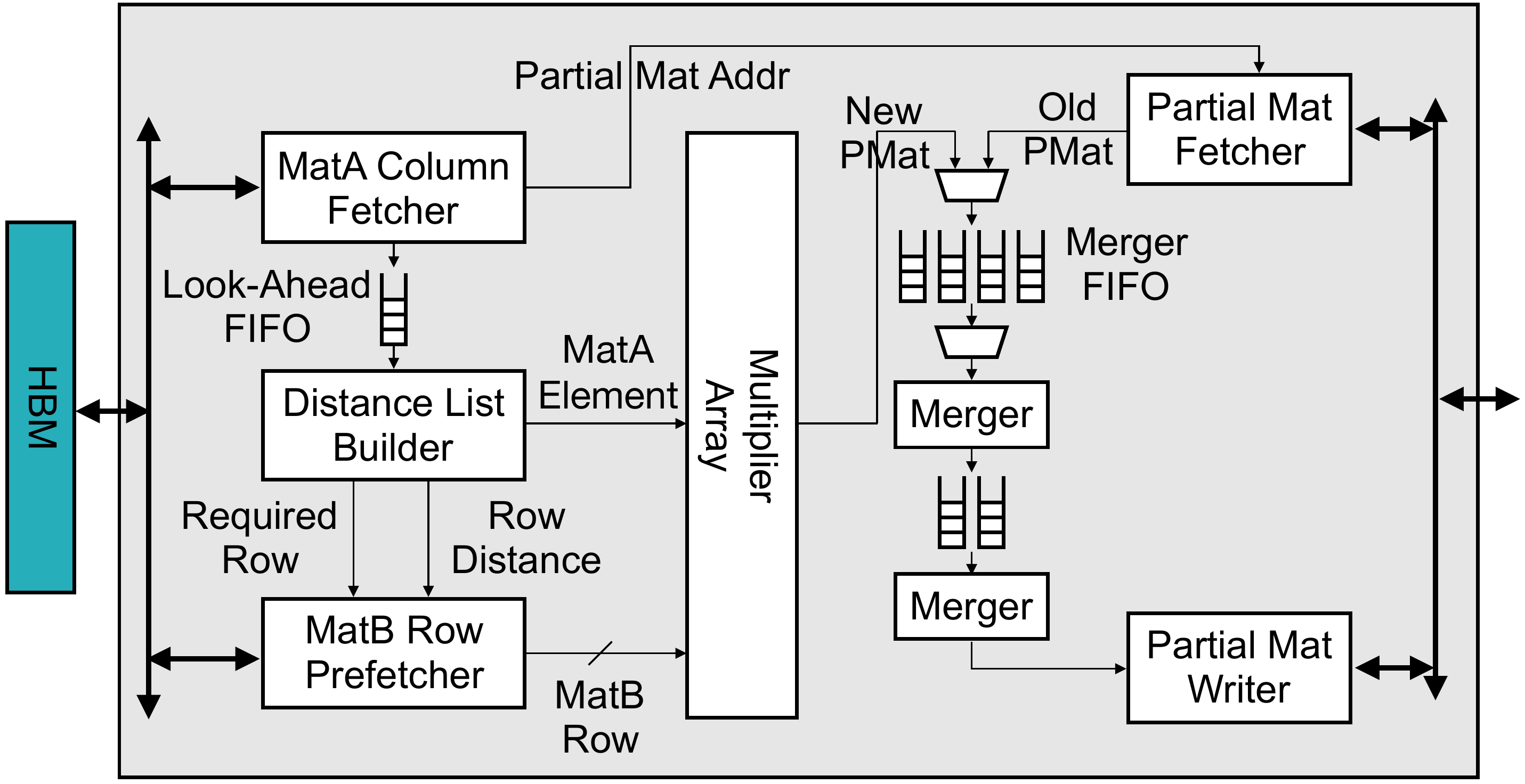}
    \vspace{-10pt}
    \caption{System architecture of proposed \name.}
    \vspace{-10pt}
    \label{fig:sysarch}
\end{figure}

Matrix condensing reduces the number of partial matrices but ruins the  input reuse of the second operand matrix since one column after condensing requires the elements from different rows of the second matrix. We solve the problem by a row prefetcher with near-optimal buffer replacement.

The prefetcher is motivated by the fact that we can predict the access order of the right matrix ahead of time. As mentioned in section~\ref{sec:condensing}, we access the left matrix from top to down. Using the column index of the left matrix, we can deduce the order of the right matrix, which can be used by the prefetcher to load data before they are consumed by the multipliers.

The prefetcher has two functions: first, fetching the data used by multipliers ahead of time to hide the latency of DRAM; second, caching the fetched data in a buffer for future reuse to reduce the amount of DRAM reading.

The first function is achieved by multiple data fetchers. We use a data fetcher for each DRAM channel. Accesses to different DRAM channels and banks are overlapped; thus, the DRAM latency can be hidden.

The second function is achieved by an on-chip buffer, which stores the rows of the right matrix. When new data is fetched, using the predicted access order, we can replace the line with the furthest next use, which can achieve near-optimal reuse if the prediction is accurate.

In Figure~\ref{fig:prefetcher}, we show a simplified example. The left column is one condensed column from the condensed first matrix. Each element is marked with its original column. The first three are original column 1, 0, and 2. On the top, we show row 0 to row 4 of the second matrix and divide them according to the length of the buffer line. The central part of the figure shows the buffer at different time steps. The red frames highlight the new contents after each time step. 

In time step 0, we load Row 1. In time step 1, we load Row 0. In time step 2, we firstly store R2-0. Then we need to spill some buffer lines to continue to load R2. We spill R0 first because R0 will be used in 7 time steps later, and R1 will be used in 3 time steps later. However, after spilling all R0 lines, R2 still has remaining parts. So we have to spill R1. Spilling a row line by line instead of as a whole can bring benefits. For example, from the time step 7 to time step 8, we only need to load R3-0 and no need for R3-1 and R3-2 because they are already in the buffer.

\subsection{System Architecture}

\label{sec:system}

The system architecture of \name is shown in Figure~\ref{fig:sysarch}.
The left matrix A is stored in CSR and fetched by condensed column, and the right matrix B is stored in CSR format in HBM.
The MatA Column Fetcher receives control instructions from the software scheduler, calculates the addresses of data in the selected columns, and fetches the elements from the left matrix.
Then the fetched elements will be sent to a look-ahead FIFO. The Distance List Builder will process the look-ahead FIFO and calculates the next use time of each row. The row index and next use time are provided to MatB Row Prefetcher to prefetch rows and decide which buffer line to spill, as illustrated in Section~\ref{sec:prefetcher}. 
To perform the associative search, we use a hash table to map row indexes to positions in the buffer and a reduction tree of next use time to decide which line to spill. As the width of the hash table is much lower than the buffer itself, and the next use time changes very few (at most one line per cycle), the power consumption of the logic is minimal.

After prefetching rows, the multiplier array will conduct the outer product and generate partial matrices in the COO format and send them to the merge tree. There is a MUX before each lowest level FIFO of the merge tree and can be configured to accept data either from the Multiplier Array or the Partial Matrix Fetcher. 
If the MUXs before each lowest level FIFOs are configured to Partial Matrix Fetcher, the scheduler will send the address of corresponding partially merged results. It will fetch the requested matrix once the FIFO is near empty.
The output of the merge tree will be connected to the Partial Matrix Writer. It buffers partial matrices before they are written to DRAM and also converts the internal COO format to the CSR format if it is the final result.
\\

\section{Evaluation}
\label{sec:experimental}

\subsection{Methodology}

\begin{table*}[t]
\centering
\begin{tabular}{l|l}

\toprule

 \multirow{2}{*}{Array Merger} & 16\x16 hierarchical merger (4\x4 top level + 4\x4 low level) with 64-bit index (32 bits for row and 32 bits for column) \\
 & and 64-bit value, running under 1GHz clock frequency.  \\ \midrule
 Merge Tree & 6 layers of array merger provides the ability of merging at most 64 arrays simultaneously. \\ \midrule
 Multiplier & 2 groups, each consists of 8 double precision floating point multipliers. \\ \midrule
 MatA Column Fetcher & Looks ahead buffer of 8192 elements. 64 fetchers support 64 columns of left matrix. \\ \midrule
 \multirow{2}{*}{MatB Row Prefetcher} & A prefetch buffer of 1024 lines\x 48 elements per line\x 12 bytes per element. 16 fetchers that load data from 16 DRAM \\
 & channels. Each can prefetch up to 48 rows before used. \\ \midrule
 Partial Matrix Fetcher & Support fetching up to 64 partial results. \\ \midrule
 Partial Matrix Writer & A FIFO of 1024 elements before they are stored into DRAM. \\ \midrule
 Main Memory & 16\x64-bit HBM channels, each channel provides 8GB/s bandwidth. \\
 \bottomrule

\end{tabular}
\caption{Architectural Setup of \name.}
\vspace{-16pt}
\label{tab:setup}
\end{table*}

\begin{center}
\begin{table}[t]
\centering
\begin{tabular}{l|c|c}
\toprule

& \name & \outerspace \\
\midrule
\midrule

Technology & 40nm & 32nm \\
\midrule
Area & 28.49 $mm^2$ & 87 $mm^2$ \\
\midrule
Power & 9.26W & 12.39W \\
\midrule
DRAM & HBM@128GB/s & HBM@128GB/s \\
\midrule
Bandwidth & \multirow{2}{*}{68.6\%} & \multirow{2}{*}{48.3\%}\\
Utilization & &\\
\bottomrule

\end{tabular}
\vspace{3pt}
\caption{Comparison with prior state-of-the-art \outerspace on area, power, and memory bandwidth utilization.}
\vspace{-18pt}
\label{tab:asicsum}
\end{table}
\end{center}
To evaluate the performance of our design, we built a cycle-accurate simulator in C++ to model the exact behavior of the hardware. Each module is abstracted as a class with a \textit{clock update} method updating the internal state of this module in each cycle, and a \textit{clock apply} method, which simulates the flip-flops in the circuit to make sure signals are updated correctly. The parameters during the simulation are listed in Table~\ref{tab:setup}.

To measure the area and power consumption, we model all the logic on the data path, which consists of the array mergers, arithmetic units, FIFOs, row prefetcher, and DRAM.
We implement the array merger in Verilog and synthesize it using Synopsys Design Compiler (DC) under TSMC 40nm library. We generate the annotated toggle rate using the xsim RTL simulator with the real input data extracted from the simulator and dump it into Switching Activity Interchange Format (SAIF) to estimate the power consumption with Design Compiler.
We then use the area and power data from \cite{galal2010energy} with the number of multiplications and additions from the simulator to model the arithmetic units (floating-point multipliers and adders).
We also dumped the size, width, and the amount of reading/writing of each SRAM/FIFO from the simulator and use CACTI to estimate the energy and area of FIFOs and prefetch buffers in the circuit.
From the simulator, we also got the exact amount of DRAM read and write. We use them to estimate our DRAM power consumption according to \cite{hbm1, hbm2}.

\begin{center}
\begin{table}[t]
\centering
\begin{tabular}{l|c|c|c|c}
\toprule

& \multicolumn{2}{c|}{Energy (nJ/FLOP)} & \multicolumn{2}{c}{Area ($mm^2$)} \\
& Outer-  & \multirow{2}{*}{SpArch} & Outer- & \multirow{2}{*}{SpArch}  \\
& SPACE & & SPACE & \\
\midrule
\midrule

Computation & 3.19 & 0.26 & 49.1 & 4.1 \\
\midrule
SRAM & 0.35 & 0.34 & 37.5 & 24.4 \\
\midrule
DRAM & 1.20 & 0.29 & N/A & N/A \\
\midrule
Crossbar & 0.21 & N/A & 0.1 & N/A \\
\midrule
Overall & 4.95 & \textbf{0.89} & 86.7 & \textbf{28.5} \\
\bottomrule

\end{tabular}
\vspace{3pt}
\caption{Energy and Power Breakdown. Savings of \name are from less DRAM access and more efficient logic.}
\vspace{-15pt}
\label{tab:breakdown}
\end{table}
\end{center}
We use multiple platforms as our baselines, including Desktop CPU, GPU, mobile CPU, and the state-of-art ASIC accelerator \outerspace. For desktop CPU, we use Intel MKL~\cite{mkl} on a 6-core Intel Core-i7 5930K CPU and measure its power with \texttt{pcm-power} tool. MKL provides math routines parallelized by OpenMP for various computational workloads. For GPU, we use both \cusparse~\cite{cusparse} and \cusp~\cite{cusp1, cusp2} and run them on an NVIDIA TITAN Xp GPU. \cusparse parallelizes the computation between matrix rows and then merges the partial results of each row with a hash table. \cusp also computes matrix rows in parallel and then sorts and merges different rows. Power consumption is measured with \texttt{nvidia-smi}.
For mobile CPU, we use Armadillo library~\cite{armadillo1, armadillo2} on a 4-core ARM A53 CPU and measure its dynamic power using a powermeter. 

\begin{figure*}[t]
    \centering
    \includegraphics[width=\textwidth]{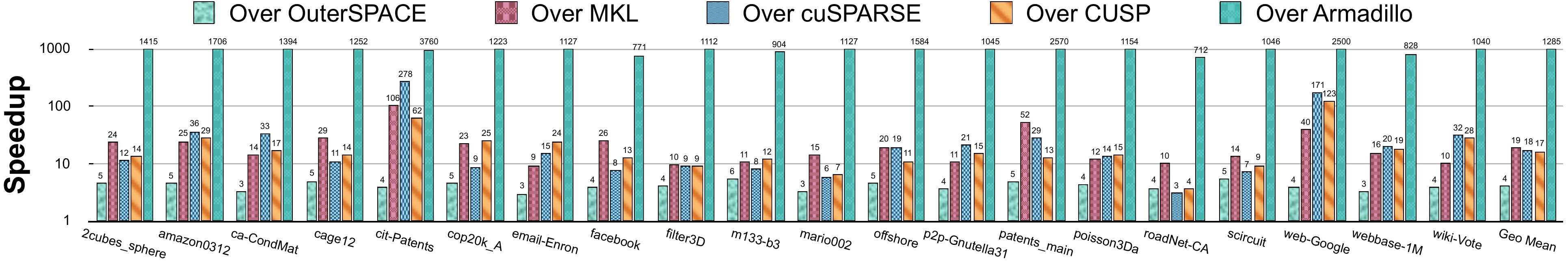}
    \vspace{-18pt}
    \caption{Speedup of \name over \outerspace, MKL, \cusparse, \cusp and ARM Armadillo on real-world matrices. We get a consistent speedup over state-of-the-art accelerator \outerspace. }
    \label{fig:perfcomp}
\end{figure*}
\begin{figure*}[t]
    \centering
    \includegraphics[width=\textwidth]{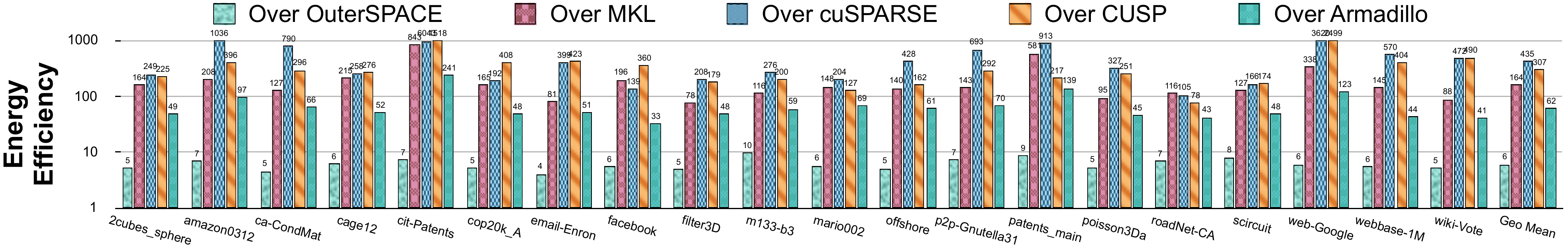}
    \vspace{-16pt}
    \caption{Energy saving of \name over \outerspace, MKL, \cusparse, \cusp and ARM Armadillo.
    }
    \label{fig:eecomp}
    \vspace{-16pt}
    
\end{figure*}
\begin{figure}[t]
    \centering
    \includegraphics[width=\columnwidth]{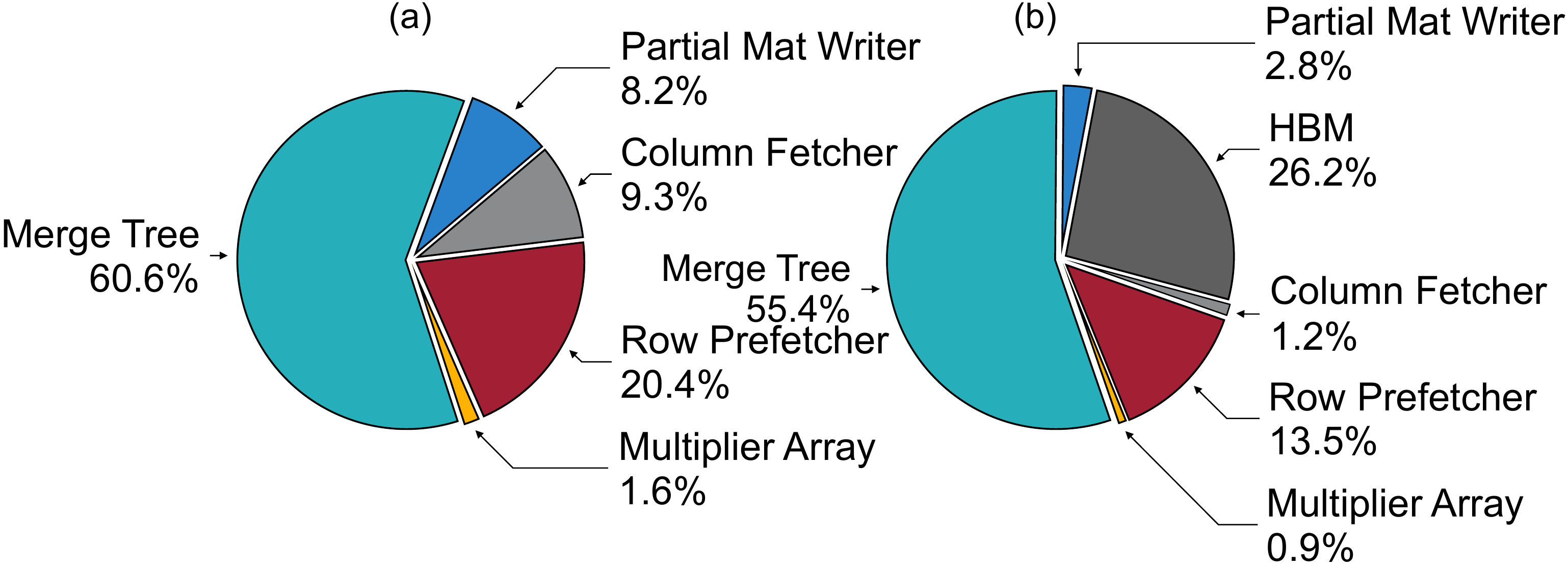}
    \caption{Area (a) and Power (b) Breakdown. The merge tree as the core of \name takes the most of area and power.}
    \label{fig:pie}
    \vspace{-15pt}
\end{figure}
In order to provide fair comparison baselines, the data type of all baselines and our architecture are \texttt{double}. For MKL, cuSPARSE, CUSP and Armadillo, we discard memory allocation and transportation time and only measure the core execution time of \spgemm functions, \ie \texttt{mkl\_sparse\_spmm}, \texttt{cusparseDcsrgemm}, \texttt{cusp::gen eralized\_spgemm} and overloaded "*" respectively. For all power measurement, we first measure the idle power of the system and then repeatedly run the workloads and measure the total power. Then the dynamic power is total power minus idle power.

\subsection{Experimental Results}

We obtain the area and power consumption under TSMC 40nm technology and compare it with state-of-the-art  SpGEMM accelerator \outerspace (Table~\ref{tab:asicsum} and Table~\ref{tab:breakdown}). For a fair comparison, we use the same DRAM power estimation as \outerspace, which is 42.6 GB/s/W. 

We also analyze the power and area per module. Figure~\ref{fig:pie} shows area and power breakdown of \name. Most of the energy and area is consumed by the merge tree, which is the core part of \name.
We evaluate the performance of \name on the SuiteSparse dataset \cite{datasetuf} and SNAP dataset \cite{datasetstan}, which are the same as \cite{pal2018outerspace}. 

Figure~\ref{fig:perfcomp} compares the \name and baselines. On average, \name is  \perfoverouterspace\x, \perfovermkl\x, \perfovercusparse\x, \perfovercusp\x \xspace and \perfoverarmadillo\x \xspace faster than \outerspace, MKL, cuSPARSE, CUSP and ARM Armadillo. 
\name outperforms the baselines on every matrix. This mainly comes from the reduction of redundant memory access of partial matrices, with 2.8\x\xspace less DRAM access compared to \outerspace. We also achieve higher memory bandwidth utilization compared to \outerspace, by virtue of hiding the DRAM latency with row prefetcher and the regular write pattern of streaming merge tree.
Figure~\ref{fig:eecomp} shows the relative energy savings. \name achieves \eeoverouterspace\x, \eeovermkl\x, \eeovercusparse\x, \eeovercusp\x \xspace and \eeoverarm\x \xspace energy saving comparing to \outerspace, MKL, cuSPARSE, CUSP and ARM Armadillo. This is because our specialized architecture designed for SpGEMM tasks removes unnecessary components like cache, instruction decoder in general processors, and uses a carefully-designed computational unit, which is more efficient on specific tasks.

\begin{figure}
    \centering
    \includegraphics[width=\columnwidth]{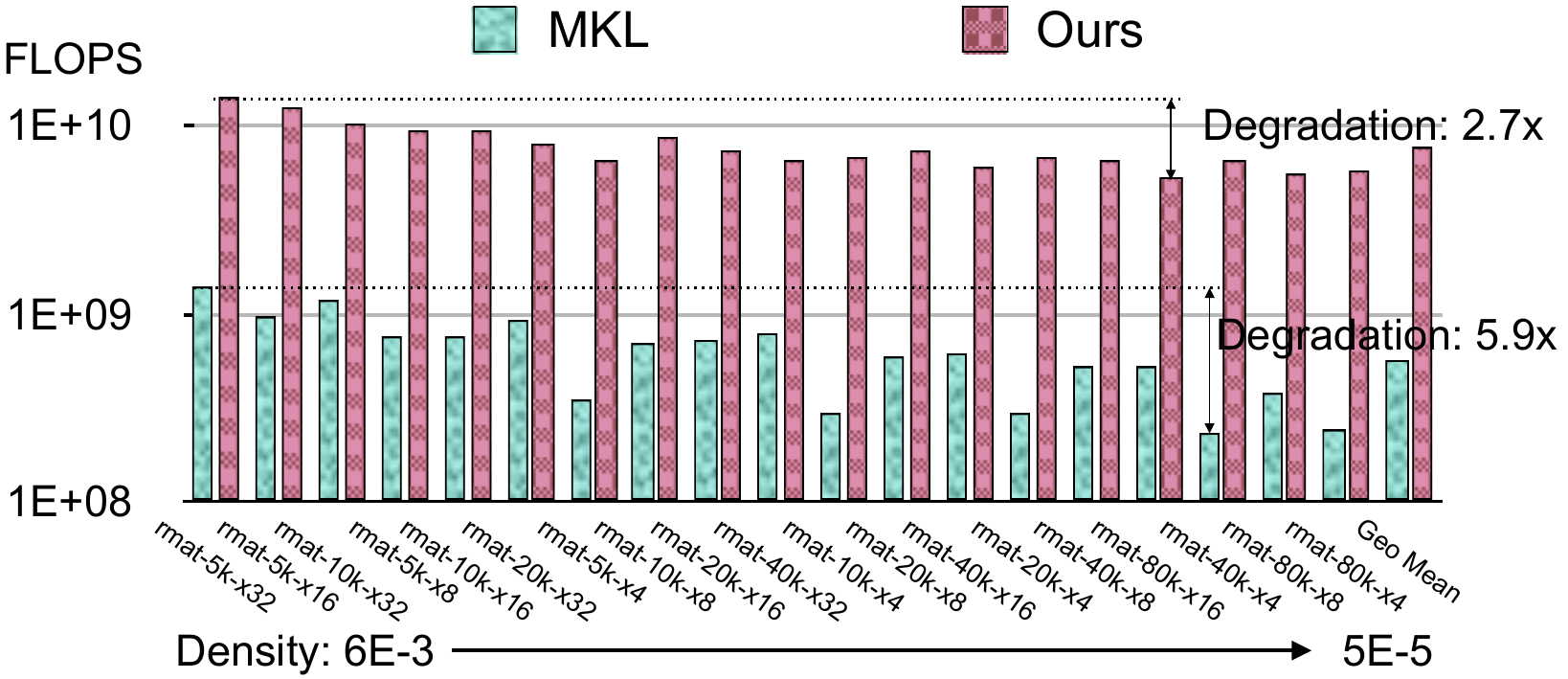}
    \vspace{-15pt}
    \caption{Performance on rMAT benchmarks. Our performance is not only higher than MKL, but also more stable when the matrices get sparser. 
    } 
    \vspace{-15pt}
    \label{fig:rmatperf}
\end{figure}
Figure~\ref{fig:rmatperf} shows the performance comparison on synthesized rMAT \cite{murphy2010introducing} data with Intel MKL. The density of matrices ranges from $6\times 10^{-3}$ to $5 \times 10^{-5}$. \name achieves over 10\x\xspace  speedup while keeping a relatively stable performance: the performance only drops by 2.7\x\xspace when matrices are sparser, while MKL suffers from larger performance degradation. The lower impact on matrix density partially derives from the outer product algorithm, which is not sensitive to the density. It also relates to our architecture design which uses a single but large merge tree rather than many small processing elements. The latter one will be more likely to suffer from the problem of load balancing, especially when matrices become sparser.
\begin{figure}
    \centering
    \includegraphics[width=0.9\columnwidth]{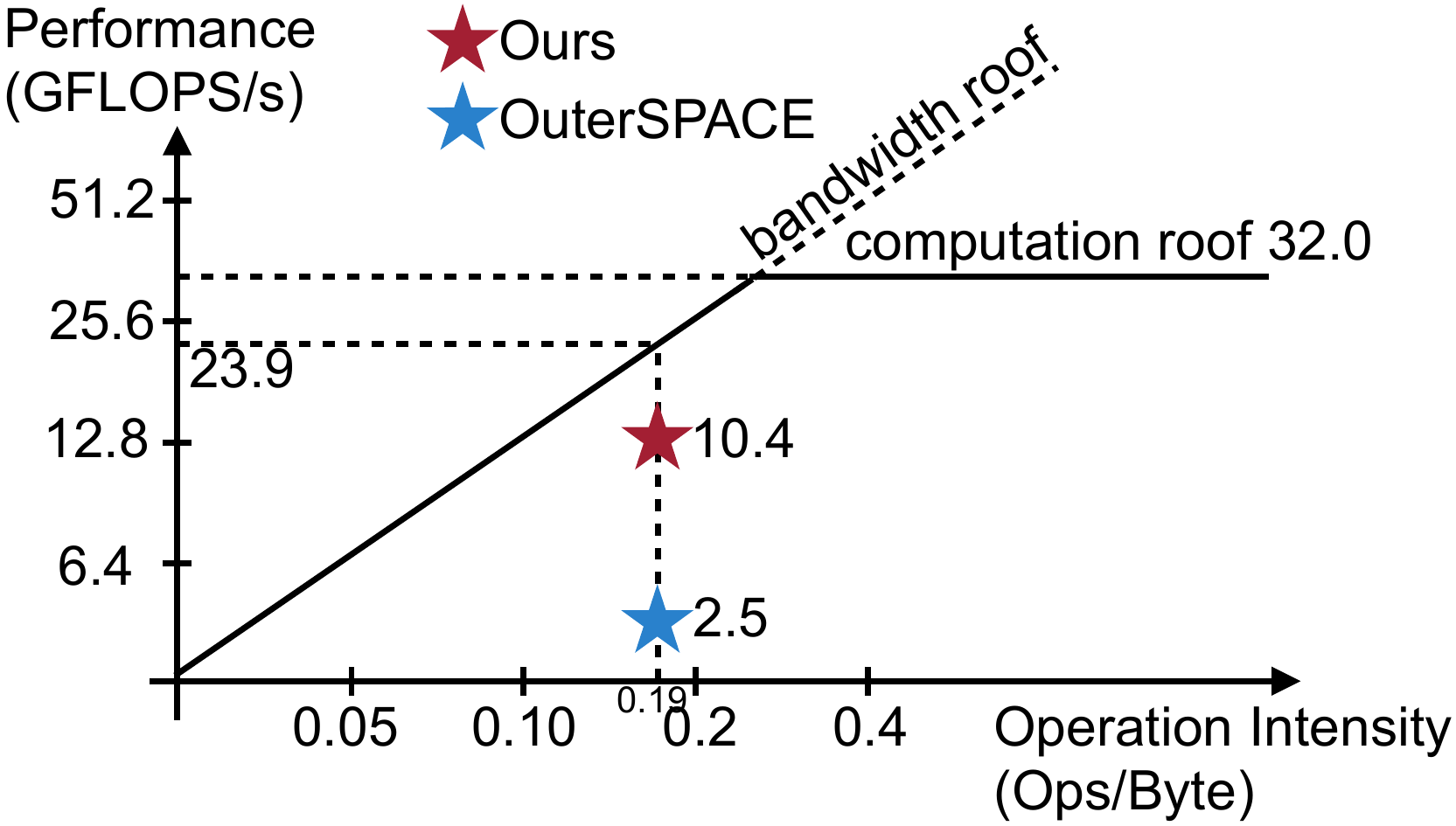}
    \vspace{-10pt}
    \caption{Roofline Model for \name and \outerspace. The theoretical operational intensity is 0.19Flops/Byte. The theoretical computation roof is 32GFlops/s. \name achieves 10.4GFlops/s while \outerspace acheives 2.5GFlops/s. Our performance is 4$\times$  higher than \outerspace by virtue of less redundant DRAM traffic for partial matrices.} 
    \label{fig:roofline}
    \vspace{-15pt}
\end{figure}

To understand how far our design is from the theoretically optimal solution, we use the roofline model to analyze our performance. Figure~\ref{fig:roofline} shows the result of roofline analysis. We calculate the theoretical operational intensity of the outer product on our dataset, which equals the number of operations divided by the size of two input matrices plus the size of merged final results, calculated to be 0.19Flops/Byte. The theoretical computation roof is 32GFlops in our design. We use 16 multipliers running in 1GHz. The peak multiplication performance is 16GFlops/s, and the overall peak performance (multiplication+addition) is 32GFlops/s. The roof in our condition is 23.9GFlops, 2.3\x\xspace over our performance, and 9.6\x\xspace over \outerspace. We are much nearer to the roof comparing to \outerspace.

\subsection{Interpreting the Performance Gain}

\begin{figure}
    \centering
    \includegraphics[width=\columnwidth]{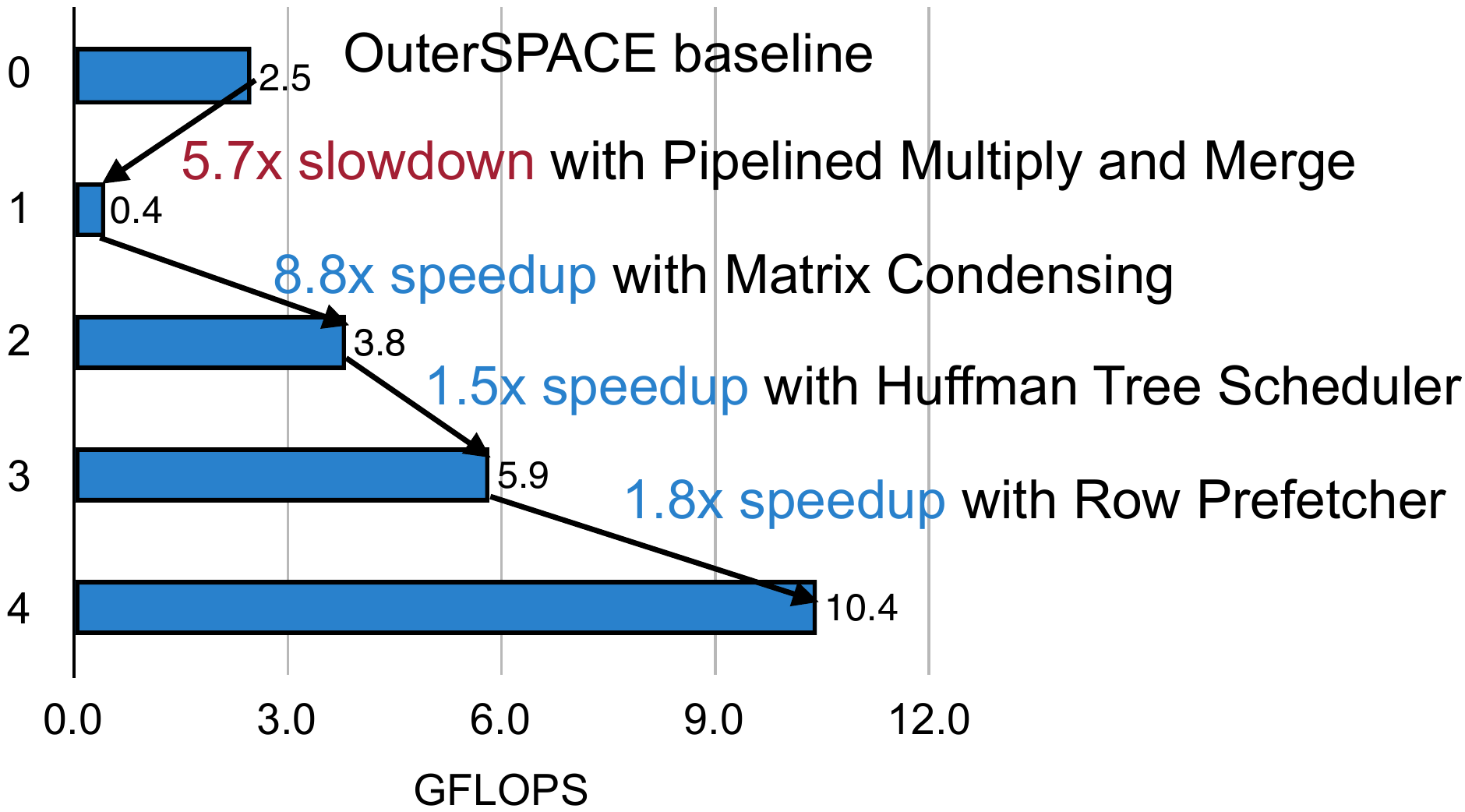}
    \vspace{-10pt}
    \caption{Dissecting the performance gain. The performance first drops due to the increased memory access of partially merged results, but it enables next three optimizations that reduce memory access more effectively: matrix condensing and Huffman tree scheduler for the output matrix, together with row prefetcher for the input matrix. Overall it achieves 4.2\x\xspace speed up over \outerspace.
    } 
    
    \label{fig:perfbreak}
    \vspace{-15pt}
\end{figure}

To analyze where our speed up comes from and demonstrate the effectiveness of several proposed ideas separately, we conduct breakdown experiments.
Figure~\ref{fig:perfbreak} shows the performance breakdown for each of our contributions. 
In the breakdown experiments, we first remove our prefetcher, scheduler, and change back to CSC/CSR matrix format but preserve the pipeline of multiply and merge phase and use a random order to select initial columns and partially merged results to merge. The performance slows down by 5.7\x \xspace comparing to \outerspace.

The DRAM read-write in OuterSPACE is mainly the intermediate results generated by the multiplier. Assuming the dimension of the matrix is $N\times N$, and there are $M$ multiplications, thus $M$ intermediate results. We also have an average of $0.5M$ final results, according to the statistics of the datasets, so the memory access is roughly $2.5M$.
In our multiply-merge pipeline, ideally, the partial matrices are not written to DRAM. However, as $N$ is much larger than the size of the merge tree (64), we still need to store lots of partially merged results to DRAM. 
Assume that the merging does not change the length of arrays, we can estimate the read times of one multiplied result $\alpha$ as follows:
\begin{align}
E &= \sum_{k=0}^{t-1} {\rm Pr}[\alpha\ {\rm is\ used\ in\ round}\ k] \\
&= \sum_{k=0}^{t-1} \frac{w}{N-k*(w-1)} \\
&= \frac{w}{w-1}*\sum_{k=0}^{t-1} \frac{1}{t+\frac{1}{w-1}-k} \\
&= \frac{w}{w-1}*\sum_{i=1}^{t} \frac{1}{\frac{1}{w-1}+i}
\end{align}
where $t=\frac{N-1}{w-1}$ is the number of rounds and $w=64$ is the size of the merge tree.
As $\frac{1}{w-1}$ is very small compared to $i$, we can ignore it:
\begin{align}
E &\approx \frac{w}{w-1}*\sum_{i=1}^{t} \frac{1}{i} \\
&\approx \frac{w}{w-1}*\ln{t}
\end{align}

On average $N\approx140,000$ in the datasets, we need to read and write the partially merged results for roughly $\ln{\frac{140000}{63}}-1\approx6.7$ times. (Minus 1 because we do not need to load/store the multiplied results in the first round). The DRAM access of partial results is roughly $6.7*2M+0.5M=13.9M$. Comparing to the $2.5M$ access of \outerspace, the 5.7\x \xspace slow down is reasonable.
We then apply matrix condensing to the left matrix, which can reduce the number of columns from 140,000 to 100, both on average. That means we can finish the merging in 2 rounds on average using a 64-way merger. However, the right matrix will be read not only once. Instead, the amount of right matrix access equals to the number of multiplications $M$. Plus the partially merged results and final results, which is roughly $((1+1/2)-1)*2M+0.5M=1.5M$ (using Formula 6), our memory access is $2.5M$, 5.5\x\xspace less than the one without condensing. 
The speedup improvement this step is 8.8\x, higher than 5.5\x \xspace because with matrix condensing, each round takes much more clock cycles than before, thus the \emph{startup overhead is reduced}, leading to higher memory utilization and higher performance.

Furthermore, we add back the Huffman Tree scheduler, which makes the memory access of partially merged results negligible. This is because the average number of non-zeroes per row is less than 64. The long columns can be scheduled near the root node in the Huffman Tree, so they will not generate partially merged results. This reduces the memory access by roughly 1.7\x \xspace (from $2.5M$ to $1.5M$), near to the actual performance we get.

We finally add back the row prefetcher. It buffers the data from the right matrix and achieves a hit rate of 62\%. Therefore, we reduce the total memory access from $1.5M$ to $0.38M + 0.5M=0.88M$, reducing DRAM access by 1.7\x, which matches the experiments.

\begin{figure}[t]
    \centering
    \includegraphics[width=\columnwidth]{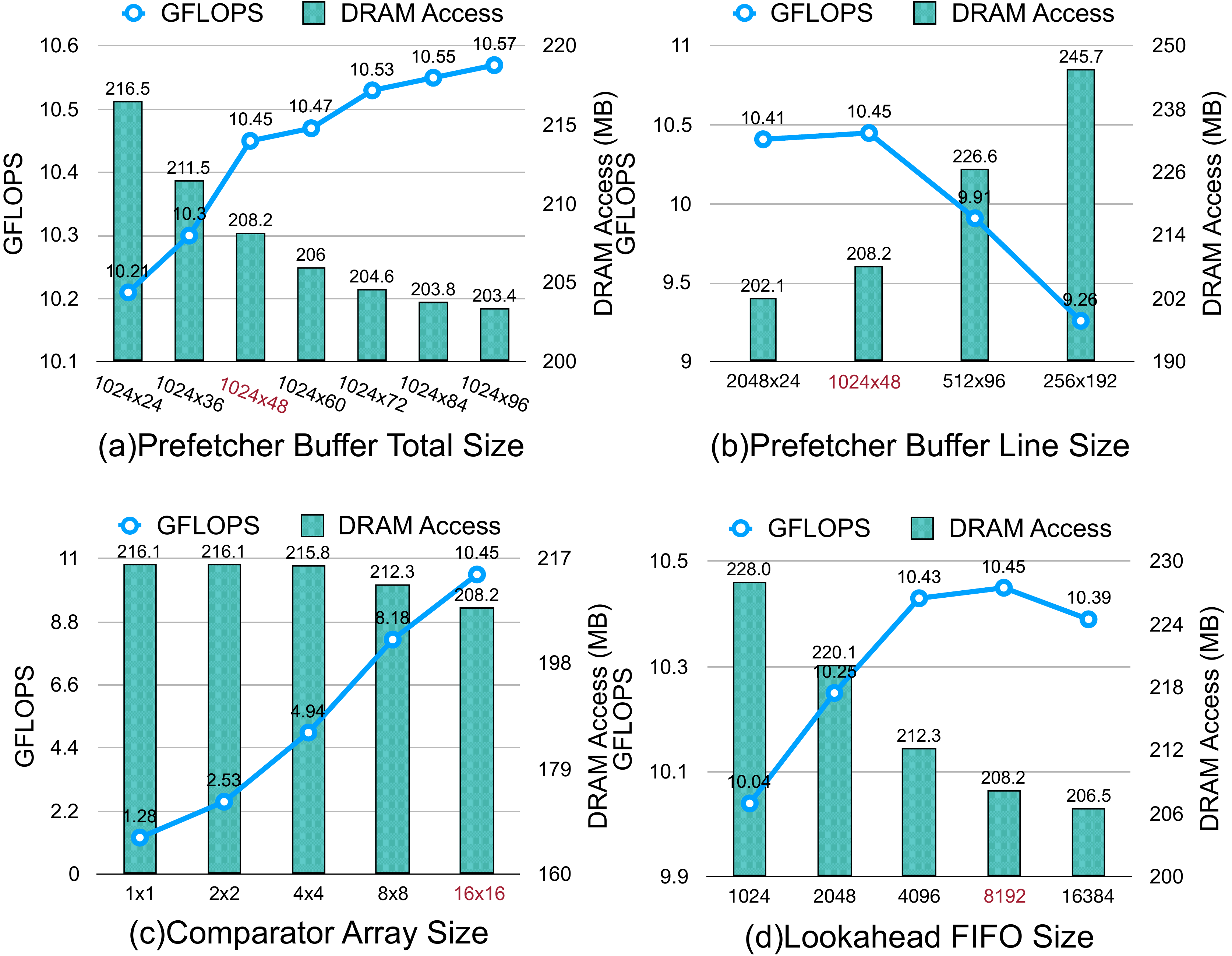}
    \caption{Design space exploration on various buffer sizes and array sizes.
    }
    \label{fig:4ab}
    \vspace{-10pt}
\end{figure}

\subsection{Design Space Exploration}

The high performance of \name comes from two factors: higher memory bandwidth utilization and lower memory access. To achieve higher bandwidth utilization, we need to make sure we are not bounded by the performance of computational units - mainly the comparator arrays in the merge tree. To achieve lower memory access, we need to make sure high reuse of the right matrix, which is mainly affected by the prefetch buffer and the look-ahead FIFO; and low partial merged results access, which is mainly determined by the size of the merge tree. 
However, higher performance usually comes with the cost of a larger area and power. We need to balance the performance and power/area carefully. We conduct design space exploration to find the optimal parameters for our design.

Figure~\ref{fig:4ab} (a)(b) shows the relationship between performance and the size of prefetch buffer ($m$\x$n$ means there are $m$ lines and each line contains $n$ elements). When the number of lines is fixed as, the longer line results in better performance, but the area will increase linearly. The difference between 1024\x60 and 1024\x48 is much smaller than the one between 1024\x48 and 1024\x36, so we choose 48 as our line size. As shown in Figure~\ref{fig:4ab} (b), When the overall buffer size is fixed as $1024\times 48=49152$, more lines leads to less memory access. However, when there are more than 1024 lines, such as 2048 lines, the benefit from less DRAM access is minor, but the increased latency of replacement logic has a greater impact on the performance, leading to smaller overall Flops/s. Thus we choose 1024 as the number of lines.

\begin{figure}[t]
    \centering
    \includegraphics[width=0.6\columnwidth]{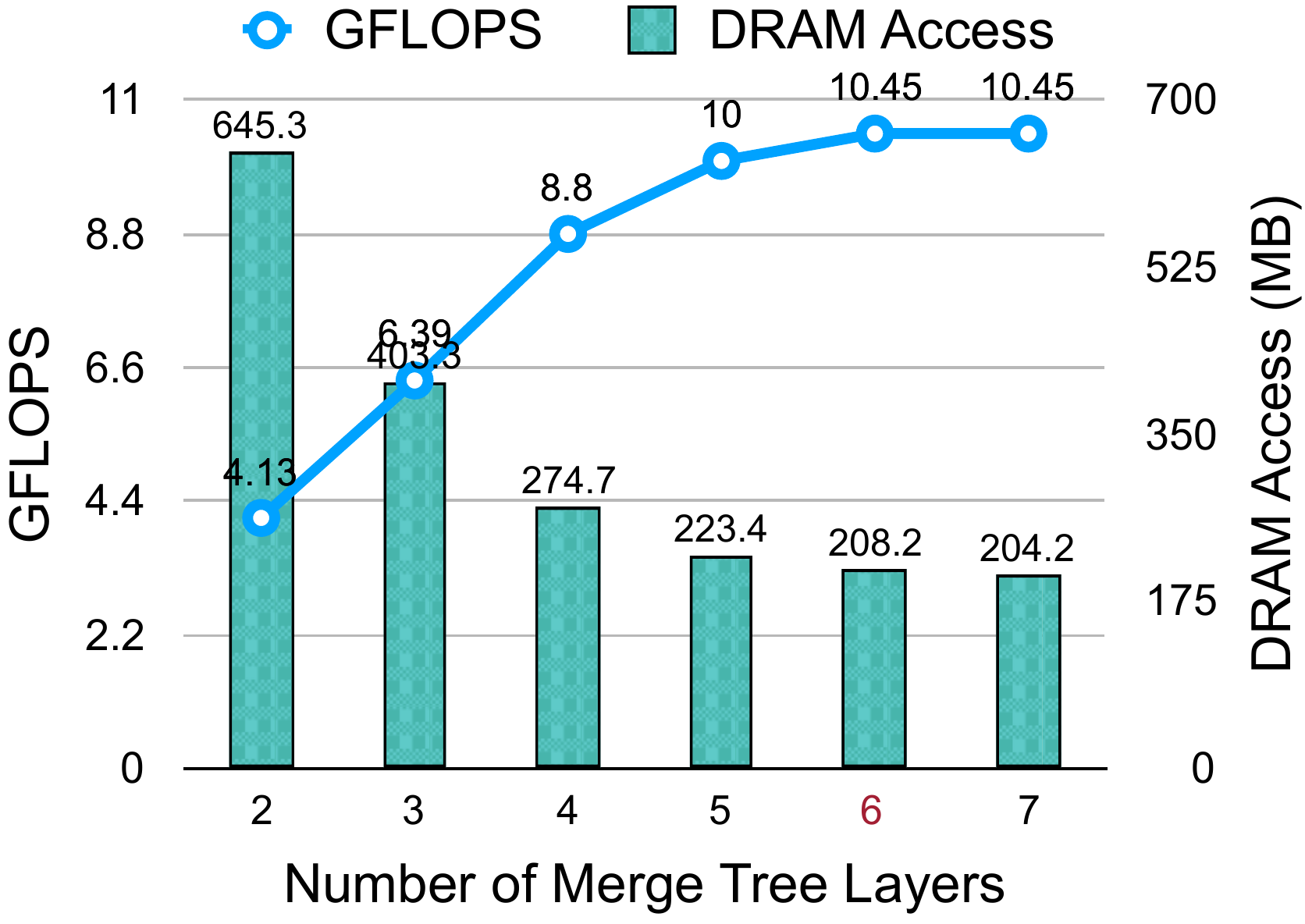}
    \vspace{-10pt}

    \caption{Design space exploration on the merge tree size. 6 layers are enough to achieve good performance, above which gives diminishing improvement.}
  
    \label{fig:1ab}
    \vspace{-10pt}

\end{figure}
Figure~\ref{fig:4ab} (c) shows the relationship between the size of comparator arrays in our merge tree and the overall performance. When the array size is less than 8\x8, the performance increases linearly. That means we are computationally bounded. When the size increases to 16\x16, the performance gain is lower, and we are bounded by memory bandwidth. So we choose 16\x16 as our comparator size, which consumes the memory bandwidth as much as possible.

Figure~\ref{fig:4ab} (d) explores the size of look-ahead FIFO. Larger FIFO makes the replacement of prefetch buffer more accurate and improves the data reuse. However, we need more time to fill the larger FIFO at the start of each round. Excessive size can lead to performance degradation. We choose 8192, which results in the highest performance, as our configuration of the look-ahead FIFO.

Figure~\ref{fig:1ab} helps us determine the size of the merge tree. Larger merge tree can process more columns at the same time but requires more FIFOs and comparator arrays. A merge tree of 6 layers and 64 ports is good enough, and the larger one does not contribute to the speedup. Therefore, we choose 6 layers to achieve the best performance without wasting area and power.

\section{Related Work}
\label{sec:related}

\textbf{Sparse Matrix-Matrix Multiplication Acceleration.}
Besides the ASIC Sparse Matrix-Matrix Multiplication accelerator \outerspace \cite{pal2018outerspace} that we mainly discussed in this paper, there are some previous explorations on SpMM on FPGA, such as \cite{lin2013design}. Contrary to our fully specialized merge tree, both of them utilized general-purpose PEs and not fully specialized for the sparse data.

There are also many works focusing on SpMM on general-purpose platforms like CPUs and GPUs.
The algorithms are mostly based on the multiply-and-insert-like Gustavson's algorithm~\cite{gustavson1978two}, which has many variants that are different in the method of inserting an element to the output matrix. We list several examples as below.
cuSPARSE~\cite{cusparse} utilizes a hash table. However, on-chip hash-table limits the scalability, and off-chip hash-table suffers from low memory utilization.
CUSP~\cite{cusp1} uses a sorting algorithm which suffers from higher complexity (sorting network) and excessive DRAM access if on-chip resources are limited.
HeapSpGEMM~\cite{azad2016exploiting} uses a heap to insert. Since the heap is hard to parallelize, the parallelism only comes from processing multiple rows simultaneously, which would suffer from the load-balance problem.
BHSPARSE~\cite{liu2014efficient} and proposed \name choose merge as the insertion method. It is less complex than sorting and remains parallelizable. It can also process the whole partial matrix rather than a single row, eliminating the load-balance issue. Compared to BHSPARSE, our design further reduces the DRAM access thanks to the specialized multiply-and-merge pipeline, which can hardly be implemented on general-purpose platforms.

\textbf{Hardware Accelerator for Merge and Intersection.}
\cite{fort2017intersecting} designed an efficient parallel GPU-based method to solve the problem of intersecting two unsorted sets. \cite{zhang2011fast} used a special data structure called \emph{bloom filter} to parallelize the intersections on GPU. \cite{ao2011efficient} implemented parallel lists intersection and index compression on GPU to speed up web search.
\cite{Han:2018:SUS:3183713.3196924} used SIMD techniques to enable a high level of parallelism in the set intersection.
Extensor~\cite{extensor} also proposes to accelerate different sparse linear algebra by hierarchically eliminating computations to the intersection operation.

\textbf{Sparse Linear Algebra.}
A number of designs are presented to accelerate sparse linear algebra on FPGA platforms. \cite{Zhuo:2005:SMM:1046192.1046202} introduced a tree of binary operators to increase the energy efficiency of SpMV on FPGAs. \cite{jamro2015algorithms} proved that separating indices comparison and computing operations could increase the throughput. The index comparison and computing in our \name system design are also separated. \cite{zou2013high} proposed a new sparse matrix storage method called "BVCSR" to compress the indices of non-zero elements, thus increasing the valid bandwidth of FPGA. \cite{elkurdi2008fpga} and \cite{grigoracs2016optimising} proposed an architecture for large-scale SpMV in the FEM problem. \cite{elkurdi2008fpga} co-designed an FPGA SpMV architecture with a matrix stripping and partitioning algorithms that enable the architecture to process arbitrarily large matrices without changing the PE quantities.

For sparse operations in deep learning algorithms, \cite{han2016eie}  proposed a specialized architecture that can leverage not only the static sparse weights but also the dynamic sparse activations. \cite{han2017ese} further extended the efficient SpMV architecture to LSTM\cite{hochreiter1997long}, a kind of recurrent neural networks. Besides SpMV, SCNN~\cite{parashar2017isca_scnn} proposed an architecture for sparse convolution operations, and SparTen~\cite{gondimalla2019micro_sparten} further solved the overhead and load-balance issue in SCNN. Instead of designing a pure hardware architecture, \cite{zhu2019micro_sparsetensorcore} focused on the algorithm and hardware co-design.
Different from those works, \name is based on the outer product algorithm, trying to maximize both input and output data reuse.

\textbf{Comparison between Different Hardware.}
The comparison of performance and energy efficiency of different platforms has been investigated by many researchers.  \cite{asano2009performance}, \cite{cong2018understanding} and \cite{cullinan2013computing} conducted comparisons by running the same benchmarks on different platforms and measuring the performance. We use similar methodology in the evaluation of \name.

\textbf{Systolic Array.}
Systolic Array \cite{kung1982systolic} is a homogeneous architecture of coupled processing units. Although the proposed comparator array shares a similar spatial architecture with systolic arrays, there exists a fundamental difference between them: the comparator array \emph{does not have long data dependencies}. Each comparator block only depends on the comparison results from the top and left neighbor block, which is computed based on the input array and not on other comparator blocks. Therefore, all the outputs of the comparator matrix can be calculated in \emph{one cycle}.

\section{Conclusion}
\label{sec:conclusion}

In this paper, we propose \name optimizing the data reuse of both the input matrix and the output matrix to accelerate sparse matrix-matrix multiplication (SpGEMM). We design a streaming-based merger to pipeline the multiply and merge phase of the outer product on-chip. We then propose a condensed matrix representation to reduce the partial matrices generated by the merge phase. We further develop a Huffman tree scheduler to make our merger scalable to larger or power-law matrices, also to reduce the memory access of partially merged results. Finally, we use a row prefetcher that achieves near-optimal buffer replacement to reduce the read of the input right matrix.
These improvements are demonstrated to be valid and fruitful. When evaluated on 20 real-world benchmarks, the memory access reduces by \dramaccesssaving\x, the performance increases by \perfoverouterspace\x, and the energy efficiency increases by \eeoverouterspace\x\xspace compared to the previous state-of-the-art. We also provide a detailed breakdown to interpret the saving of each step, offering insights to the specialized accelerator designs.

\section*{Acknowledgement}

Part of this work was supported under an NSF HDR Award \#1934700 and DARPA SDH program. We thank MIT Data Science and AI Lab (DSAIL) for supporting this research. We thank Joel Emer, Saman Amarasinghe, Julian Shun, Stephen Keckler, and Christopher Fletcher for inspiring discussions. This research was, in part, funded by the U.S. Government. The views and conclusions contained in this document are those of the authors and should not be interpreted as representing the official policies, either expressed or implied, of the U.S. Government.

\bibliographystyle{ieeetr}
\bibliography{main.bib}

\end{document}